\documentclass[12pt]{article}

\makeatletter

\RequirePackage{cite}
\RequirePackage{times}
\RequirePackage{ifthen}

\advance \topmargin by -\headheight
\advance \topmargin by -\headsep
     
\evensidemargin \oddsidemargin
\renewcommand\@biblabel[1]{#1.}

\def\@cite#1#2{$^{\mbox{\scriptsize #1\if@tempswa , #2\fi}}$}

\newcommand{\spacing}[1]{\renewcommand{\baselinestretch}{#1}\large\normalsize}
\spacing{1}

\def\@maketitle{%
  \newpage\spacing{1}\setlength{\parskip}{12pt}%
    {\Large\bfseries\noindent\sloppy \textsf{\@title} \par}%
    {\noindent\sloppy \@author}%
}

\newenvironment{affiliations}{%
    \setcounter{enumi}{1}%
    \setlength{\parindent}{0in}%
    \slshape\sloppy%
    \begin{list}{\upshape$^{\arabic{enumi}}$}{%
        \usecounter{enumi}%
        \setlength{\leftmargin}{0in}%
        \setlength{\topsep}{0in}%
        \setlength{\labelsep}{0in}%
        \setlength{\labelwidth}{0in}%
        \setlength{\listparindent}{0in}%
        \setlength{\itemsep}{0ex}%
        \setlength{\parsep}{0in}%
        }
    }{\end{list}\par\vspace{12pt}}

\renewenvironment{abstract}{%
    \setlength{\parindent}{0in}%
    \setlength{\parskip}{0in}%
    \bfseries%
    }{\par\vspace{-6pt}}

\renewcommand{\section}{\@startsection {section}{1}{0pt}%
    {-6pt}{1pt}%
    {\bfseries}%
    }
\renewcommand{\subsection}{\@startsection {subsection}{2}{0pt}%
    {-0pt}{-0.5em}%
    {\bfseries}*%
    }

\newenvironment{methods}{%
    \section*{Methods}%
    \setlength{\parskip}{12pt}%
    }{}

\newenvironment{addendum}{%
    \setlength{\parindent}{0in}%
    \small%
    \begin{list}{Acknowledgements}{%
        \setlength{\leftmargin}{0in}%
        \setlength{\listparindent}{0in}%
        \setlength{\labelsep}{0em}%
        \setlength{\labelwidth}{0in}%
        \setlength{\itemsep}{12pt}%
        }
    }
    {\end{list}\normalsize}

\usepackage{float}
\usepackage{commath}
\usepackage{amsmath}
\usepackage{amssymb}
\usepackage{bm}
\usepackage{graphicx}
\usepackage[symbol]{footmisc}
\usepackage{rotating}
\usepackage[para]{threeparttable}

\newcommand\ngc{NGC~3923 }%
\newcommand\msol{$M_{\odot}$}%
\def\simgt{\lower.5ex\hbox{$\; \buildrel > \over \sim \;$}}%
\def\simlt{\lower.5ex\hbox{$\; \buildrel < \over \sim \;$}}%
\def\tas{Type Ia SN}%
\def\tase{Type Ia SNe}%
\def\sn1a{SNe Ia}%

\def\uname{SN~2018aoz}

\def\he1{${\rm{^{1}He}}$}%
\def\ni56{${\rm{^{56}Ni}}$}%
\def\co56{${\rm{^{56}Co}}$}%
\def\fer56{${\rm{^{56}Fe}}$}%
\def\fe52{${\rm{^{52}Fe}}$}%
\def\chr48{${\rm{^{48}Cr}}$}%
\def\ti44{${\rm{^{44}Ti}}$}%
\def\dm15{$\Delta$$M_{15}(B)$}%

\def\sbv{$s_{BV}$}%
\def\tp{$t_{\rm p}$}%
\def\t0{$t_{\rm 0}$}%

\def\degr{$\hbox{$^{\circ}$ }$}
\def\arcmin{$\hbox{$^{\prime}$ }$}
\def\farcs{$\hbox{$^{\prime\prime}$}$}
\def\chisqr{$\chi^2_{\rm R}$}%

\title{Infant-phase reddening by surface Fe-peak elements in a normal Type Ia Supernova}

\author{Yuan Qi Ni$^1$, $^{\ast}$Dae-Sik Moon$^1$, Maria R. Drout$^{1}$, Abigail Polin$^{2,3}$, David J. Sand$^4$, Santiago Gonz\'alez-Gait\'an$^5$, Sang Chul Kim$^{6,7}$, Youngdae Lee$^{8,6}$, Hong Soo Park$^{6,7}$, D. Andrew Howell$^{9,10}$, Peter E. Nugent$^{11,12}$, Anthony L. Piro$^{2}$, Peter J. Brown$^{13,14}$,  Llu\'is Galbany$^{15,16}$, Jamison Burke$^{10,9}$,  Daichi Hiramatsu$^{10,9,17,18}$, Griffin Hosseinzadeh$^{4}$, Stefano Valenti$^{19}$, Niloufar Afsariardchi$^1$, Jennifer E. Andrews$^4$, John Antoniadis$^{20,21,22}$, Iair Arcavi$^{23,24}$, Rachael L. Beaton$^{39,40,25,2}$,  K. Azalee Bostroem$^{26}$, Raymond G. Carlberg$^{1}$, S. Bradley Cenko$^{27,28}$, Sang-Mok Cha$^{6,29}$, Yize Dong$^{19}$, Avishay Gal-Yam$^{30}$, Joshua Haislip$^{31}$, Thomas W.-S. Holoien$^{2}$,  Sean D. Johnson$^{25}$, Vladimir Kouprianov$^{31,32}$, Yongseok Lee$^{6,29}$, Christopher D. Matzner$^1$, Nidia Morrell$^{33}$, Curtis McCully$^{10,9}$, Giuliano Pignata$^{34,35}$, Daniel E. Reichart$^{31}$, Jeffrey Rich$^{2}$, Stuart D. Ryder$^{36,37}$, Nathan Smith$^4$, Samuel Wyatt$^4$, Sheng Yang$^{38}$}

\begin{document}

\maketitle

\begin{affiliations}
\item{David A. Dunlap Department of Astronomy and Astrophysics, University of Toronto, Toronto, ON, Canada}
\item{The Observatories of the Carnegie Institution for Science, Pasadena, CA, USA}
\item{TAPIR, Walter Burke Institute for Theoretical Physics, Caltech, Pasadena, CA, USA}
\item{Department of Astronomy/Steward Observatory, University of Arizona, Tucson, AZ, USA}
\item{CENTRA-Centro de Astrof\'{i}sica e Gravita\c{c}\~ao and Departamento de F\'{i}sica, Instituto Superior T\'ecnico, Universidade de Lisboa, Avenida Rovisco Pais, Lisboa, Portugal}
\item{Korea Astronomy and Space Science Institute, Daejeon, Republic of Korea}
\item{Korea University of Science and Technology, Daejeon, Republic of Korea}
\item{Department of Astronomy and Space Science, Chungnam National University, Daejeon, Republic of Korea}
\item{Department of Physics, University of California, Santa Barbara, CA, USA}
\item{Las Cumbres Observatory, Goleta, CA, USA}
\item{Department of Physics and Astronomy, Texas A\&M University, College Station, TX, USA}
\item{Lawrence Berkeley National Laboratory, Berkeley, CA, USA}
\item{Department of Astronomy, University of California, Berkeley, CA, USA}
\item{George P. and Cynthia Woods Mitchell Institute for Fundamental Physics \& Astronomy, College Station, TX, USA}
\item{Institute of Space Sciences (ICE, CSIC), Campus Universitat Autònoma de Barcelona, Carrer de Can Magrans, Barcelona, Spain}
\item{Institut d’Estudis Espacials de Catalunya (IEEC), Barcelona, Spain.}
\item{Center for Astrophysics \textbar{} Harvard \& Smithsonian, Cambridge, MA, USA}
\item{The NSF AI Institute for Artificial Intelligence and Fundamental Interactions}
\item{Department of Physics and Astronomy, University of California, Davis, CA, USA}
\item{Institute of Astrophysics, FORTH, Department of Physics, University of Crete, Voutes, University Campus, Heraklion, Greece}
\item{Max-Planck-Institut f\"ur Radioastronomie, Bonn, DE}
\item{Argelander Institut f\"ur Astronomie, Bonn, DE}
\item{The School of Physics and Astronomy, Tel Aviv University, Tel Aviv, Israel}
\item{Canadian Institute for Advanced Research (CIFAR), Toronto, Ontario, Cana}
\item{Department of Astrophysical Sciences, Princeton University, Princeton, NJ, USA}
\item{DIRAC Institute, Department of Astronomy, University of Washington, Seattle, WA, USA}
\item{Astrophysics Science Division, NASA Goddard Space Flight Center, Greenbelt, MD, USA}
\item{Joint Space-Science Institute, University of Maryland, College Park, MD, USA}
\item{School of Space Research, Kyung Hee University, Yongin, Republic of Korea}
\item{Department of Particle Physics and Astrophysics, Weizmann Institute of Science, Rehovot, Israel}
\item{Department of Physics and Astronomy, University of North Carolina at Chapel Hill, Chapel Hill, NC, USA}
\item{Central (Pulkovo) Observatory of Russian Academy of Sciences, Saint Petersburg, Russia}
\item{Las Campanas Observatory, Carnegie Observatories, La Serena, Chile}
\item{Departamento de Ciencias Fisicas, Universidad Andres Bello, Santiago, Chile}
\item{Millennium Institute of Astrophysics (MAS), Providencia, Santiago, Chile}
\item{Department of Physics and Astronomy, Macquarie University, NSW, Australia}
\item{Macquarie University Research Centre for Astronomy, Astrophysics \& Astrophotonics, Sydney, NSW, Australia}
\item{The Oskar Klein Centre, Department of Astronomy, Stockholm University, AlbaNova, Stockholm, Sweden}
\item{Hubble Fellow: Rachael L. Beaton}
\item{Carnegie-Princeton Fellow: Rachael L. Beaton}
\item[*] Correspondence should be addressed to D.-S.M.~(email: moon@astro.utoronto.ca).
\end{affiliations}

\begin{abstract}
Type Ia Supernovae are thermonuclear explosions of white dwarf stars. 
They play a central role in the chemical evolution of the Universe
and are an important measure of cosmological distances. However, outstanding questions
remain about their origins.
Despite extensive efforts to obtain natal information from their earliest signals, observations have thus far failed to identify how the majority of them explode. 
Here, we present infant-phase detections of \uname\ from a brightness of $-$10.5 absolute AB magnitudes---the lowest luminosity early Type Ia signals ever detected---revealing a hitherto unseen plateau in the $B$-band that results in a rapid redward color evolution between 1.0 and 12.4 hours after the estimated epoch of first light.
The missing $B$-band flux is best-explained by line-blanket absorption from Fe-peak elements in the outer 1\% of the ejected mass. The observed $B-V$ color evolution of the SN also matches the prediction from an over-density of Fe-peak elements in the same outer 1\% of the ejected mass, whereas bluer colors are expected from a purely monotonic distribution of Fe-peak elements.
The presence of excess nucleosynthetic material in the extreme outer layers of the ejecta points to enhanced surface nuclear burning or extended sub-sonic mixing processes in some normal \tas\ explosions.
\end{abstract}

Type Ia Supernovae (SNe)
are the main source of iron (Fe)-peak elements in the Universe\cite{Matteucci2012book} and their use in measuring extragalactic distances led to the discovery of the accelerated cosmological expansion and dark energy\cite{Riess1998aj, perlmutter1999apj}.
Despite their fundamental importance, the explosion mechanisms of \tase\ remain a matter of extensive debate\cite{Maoz2014araa},
particularly for the ``normal'' events, comprising $\sim$ 70\% of their population\cite{Blondin2012aj}.
Normal \tase\ may be ignited by nuclear burning in a white dwarf (WD) when binary accretion or merger causes its mass to reach the critical Chandrasekhar limit ($\sim$ 1.4\,\msol).
Alternatively, recent
studies have shown that they may also arise from sub-Chandrasekhar-mass WDs via a helium-shell double-detonation (He-shell DDet), wherein the detonation of a thin helium layer on the surface of the WD subsequently ignites carbon in the core\cite{Polin2019apj, Townsley2019apj, Shen2021apjl}.
Normal \tase\ may arise from multiple explosion channels and it is uncertain what fraction of them are produced by different explosion channels. 
It has been suggested that He-shell DDets can potentially account for up to $\sim$ 40\% of them\cite{Polin2019apj}.

Historically, most \tase\ were discovered and monitored around the peak of their light curves 2--3 weeks post-explosion, when their emission is dominated by the decay of \ni56\ concentrated in the center of their ejecta.
However, multiple explosion processes, such as sub-sonic mixing\cite{Reinecke2002aa} and surface He burning\cite{Polin2019apj, Maeda2018apj}, predict over-densities of Fe-peak elements in the shallow outer layers of the ejecta, leading to excess emission and short-lived spectroscopic features in the faint and elusive ``infant'' phase within a few days post-explosion. 
Excess red emission identified between 1 and 5 days in at least two spectroscopically peculiar events
was interpreted to be from the ashes of a He-shell detonation\cite{Jiang2017nat, De2019apj, Bulla2020apj}.
On the other hand, multiple events from the normal\cite{Marion2016apj, Hosseinzadeh2017apj, Shappee2018apj, Dimitriadis2019apj} and overluminous\cite{Jiang2018apj, Stritzinger2018apj} subtypes have been observed with excess blue emission until 5 days. 
There are several possible interpretations for the blue emission, including
excess heating from a \ni56\ over-density\cite{Sand2018apj, Hosseinzadeh2017apj}.
For the vast majority of \tase\ observed between 1 and 5 days, the rising parts of their light curves match simple power-law profiles\cite{Nugent2011nat, Foley2012apj, Olling2015nat, Cartier2017mnras, Holmbo2019aa, Miller2020apj}, consistent with a centrally concentrated and monotonic distribution of \ni56\ under the ejecta surface.
Overall, observations to date have tightened constraints on the distribution of nucleosynthetic elements in normal \tase, but 
how they explode still remains uncertain\cite{Bulla2020apj}.
Deep multi-color observations covering the infant phase within $\sim$ 1 day post-explosion can shed new light on this longstanding problem.

{\noindent \bf Discovery and Follow-up Observations of SN~2018aoz}

\uname\ was first detected at 00h54m on 29 March 2018 Universal Time (UT) in a $B$-band image taken by the Korea Microlensing Telescope Network\cite{Moon2016spie, Kim2016jkas} (KMTNet)
of a field containing the elliptical galaxy NGC 3923\cite{Norris2011mnras} at redshift $z=0.00580$ (see Table~\ref{tab:data}).
Unlike previous studies of \tase, our observations were obtained in three bands ($BVI$) from the moment of first detection, at an average cadence of 4.7 hours.
The first detections in the $V$ and $I$ bands followed 2 minutes and 2.4 hours later, respectively, while it was not detected in images from the previous night.
The source was also identified by the Distance Less than 40 Mpc survey\cite{Tartaglia2018} 1.1 days after the first KMTNet detection, reported to the Transient Name Server at 07h25m on 2 April 2018 UT, and classified as a normal \tas\ with a spectrum obtained by the Las Cumbres Observatory\cite{Brown2013pasp} at 09h25m on 2 April 2018 UT.
The discovery triggered an extensive follow-up campaign, in which ultra-violet (UV), optical, and near-infrared (NIR) imaging and spectroscopy were obtained.

Figure~\ref{fig:lc} compares part of our high-cadence $UBVri$ light curves of \uname\ (see Supplementary Information for the full UV to NIR light curves)
with early light curves of other normal \tase\ after Galactic extinction correction (Methods).
Note that the KMTNet $BVI$-band observations are calibrated to the nearest American
Association of Variable Star Observers (AAVSO) standard filters ($BVi$; see Methods).
The $B$-band light curve peak absolute magnitude, $M_{max}(B)$ = $-$19.32 mag, and decline rate over the first 15 days post-peak, \dm15 = 1.12 mag---two parameters commonly used to classify Type Ia SNe\cite{Parrent2014apss}---identify \uname\ as a normal Type Ia.
The early light curves of \uname\ rise faster than most normal events such as SNe~2011fe\cite{Nugent2011nat, Pereira2013aa}, 2018oh\cite{Li2019apj, Dimitriadis2019apj}, and 2017cbv\cite{Hosseinzadeh2017apj, Sand2018apj}, which have similar $M_{max}(B)$ and \dm15\ values as \uname,
though the total rest-frame rise time of \uname\ from its onset to $B$-band maximum (15.32 days) is consistent with events near the lower extreme of the rise time distribution of the normal population (ranging in 15--22 days\cite{Miller2020apj}).

{\noindent \bf Earliest Observations: $B$-band Plateau and Excess Emission}

The onset of the light curve (or ``epoch of first light'') is estimated to be 23h54m on 28 March 2018 UT (MJD 58205.9958), or 1.0 $\pm$ 0.5 hours before the first detection (see below), making it the earliest 3-$\sigma$ detection for any \tas.
Figure~\ref{fig:lc} (inset) compares the earliest observations of \uname\ to those of other normal \tase\ in filters nearest to the $V$ band.
The observations of \uname\ over the first $\sim$ 0.5 days constitute the lowest luminosity signals ever detected from an early \tas---up to a factor of 3 below the first detected luminosity of SN~2011fe\cite{Nugent2011nat}.
In this unexplored ``infant'' phase and depth, the $B$-band fluxes are nearly constant (the ``$B$-band plateau'') while the $Vi$ bands rise rapidly at a rate of $\sim$ 2 mags/day, resulting in an abrupt redward evolution of the $B-V$ color by 1.5 mags between 1.0 and 12.4 hours after first light (Figure~\ref{fig:colorobs}).
These features have never been identified in any \tas\ to date.

In order to estimate the epoch of first light, we fit a power-law model to the rising light curves of \uname\ (Supplementary Section 1), as typically adopted in the analysis of \tase\cite{Nugent2011nat,Foley2012apj, Goobar2015apj, Olling2015nat, Cartier2017mnras, Jiang2017nat, Holmbo2019aa, Dimitriadis2019apj, Miller2020apj}.
The $B$-band light curve during 0--0.5 days was omitted from the fitting 
to avoid the uncertainty associated with the $B$-band plateau.
Figure~\ref{fig:early} (left panels) compares the best-fit power-law to the observed light curves of the SN over the fitting interval (0--7 days).
In the infant phase of 0--0.5 days, faint excess emission over the simple power-law fit is evident in the $V$ and $i$ bands, simultaneous with the $B$-band plateau.
In subsequent epochs, the power-law provides an acceptable fit to the rising light curves, consistent with the vast majority of \tase\ such as 2011fe (Figure~\ref{fig:lc}), while there is no evidence for excess emission in \uname\ at the brightness level of the excess emission observed in SNe 2018oh and 2017cbv during 1--5 days.

{\noindent \bf Origin of Infant-Phase Emission in the Extreme Outer Ejecta}

The appearance of a short-duration $B$-band plateau and simultaneous excess emission over the power-law emission almost immediately following the SN explosion would localize the mechanisms responsible for these features to the extreme outer ejecta where the optical depth is lower.
However, the short 15.3-day rise time of \uname\ compared to the 18.9-day average for normal \tase\ could, in principle, indicate a long ``dark phase'' between the epochs of explosion and first light due to the time delay for radioactive photons to diffuse out of the ejecta\cite{Piro&Nakar2013apj}.
To examine the presence of such a dark phase in \uname, we estimate the explosion epoch, $t_{\rm exp}$, by fitting its photospheric velocity evolution traced by the Si~II~$\lambda$6355~\AA\ absorption feature with a power-law model that has provided a good fit to other \tase\cite{Hosseinzadeh2017apj, Piro&Nakar2014apj} (Supplementary Section 2), obtaining $t_{\rm exp}$ = MJD 58205.6 $\pm$ 0.7 (0.4 days prior to first light).
Since this method is independent of the distribution of radioactive photons, it is also dark-phase independent.
The difference between the epochs of explosion and first light constrains the dark phase in \uname\ to be $\lesssim$ 1 day, disfavouring the dark phase explanation for its short rise time among normal events.
Instead, the shorter rise time can be attributed to a smaller effective ejecta mass along the line of sight,
or a smaller total ejecta mass assuming spherically symmetric ejecta (Supplementary Section 4).

The mass of material in the outer ejecta that is visible via photon diffusion at time $t$ post-explosion, $\Delta M(t)$, can be estimated via the following equation derived from a polytropic (n = 3) ejecta profile expected for an exploding WD undergoing homologous expansion\cite{Piro&Nakar2013apj}.
\begin{equation}
    \Delta M \approx 2\times 10^{-2} 
    \left(\frac{\kappa}{0.1\ {\rm cm}^{2}\ {\rm g}^{-1}}\right)^{-0.88} 
    \left(\frac{M_{\rm ej}}{1.4\ M_{\odot}}\right)^{-0.32} 
    \left(\frac{E_{\rm ej}}{10^{51}\ {\rm ergs}}\right)^{0.44}
    \left(\frac{t-t_{\rm exp}}{\rm 1\ day}\right)^{1.76}\ M_{\odot}
\label{eq:diff}
\end{equation}
where $M_{\rm ej}$ and $E_{\rm ej}$ are the mass and kinetic energy of the SN ejecta, respectively, and $\kappa \sim$ 0.1~cm$^2$~g$^{-1}$ is expected for \ni56-dominated opacity\cite{Pinto&Eastman2000apj}.
The equation can be re-written compactly in terms of the parameter $\tau_m$, the geometric mean of the diffusion and expansion timescales:
\begin{equation}
    \frac{\Delta M}{M_{\rm ej}} \approx 1.3 
    \left(\frac{t-t_{\rm exp}}{\tau_m}\right)^{1.76}\ M_{\odot}\,, \quad\quad\tau_m = \left( \frac{\kappa}{13.8\, c}\right)^{1/2} \left( \frac{6\,M_{\rm ej}^3}{5 \,E_{\rm ej}}\right)^{1/4}
\label{eq:taum}
\end{equation}
Adopting $\tau_m$ = 9.5 from modelling the observed bolometric luminosity of the SN (Supplementary Sections 3 and 4), this locates the origin of the infant-phase features of \uname\ visible $\lesssim$ 1 day post-explosion within the outer $\sim$ 2\% of the SN-ejected mass
along the line of sight.

{\noindent \bf Origin of the Infant-Phase $B$-band Plateau and Red $B-V$ Color}

We examine whether the infant-phase $B$-band plateau and redward $B-V$ color evolution are consistent with blackbody emission from cooling shock interactions that can be readily expected in early \tase\ (e.g., ejecta collision with a companion\cite{Kasen2010apj}).
Figure~\ref{fig:specfig} compares the observed spectral energy distribution (SED) of \uname\ in $BVi$ filters at 0.5 days, when the $B-V$ color peaks,
to those of various blackbodies. 
Clearly, it is impossible to fit all three $BVi$ bands at this epoch with a single blackbody distribution.
The $Vi$-band fluxes alone are consistent with a 13000 $\pm$ 5000~K blackbody, which is within the expectations for shock-heated SN ejecta\cite{Piro2010apj, Kasen2010apj}.
However, accommodating the low $B$-band flux level
requires either unreasonably cold temperatures for early Type Ia SNe or a break in the spectrum between the $B$ and $V$ bands ($\sim$ 5000~\AA) with $\gtrsim$ 70\% flux suppression bluewards of the break (Supplementary Section 6). 
This indicates that while the underlying emission may be thermal in origin, a \emph{suppression} of the $B$-band flux relative to the blackbody distribution is required to reproduce the observed SED of \uname\ during the infant phase.
Note that continuous extinction (e.g., by circumstellar dust) is unable to explain the suppression, even for the case of extreme dust extinction (Supplementary Section 6).
For a photosphere radius of 2.5 $\times$ 10$^{14}$~cm at 0.9 days post-explosion---estimated by adopting a polytropic (n = 3) ejecta profile\cite{Piro&Nakar2013apj} with $\tau_m$ = 9.5 and the observed Si II velocity near peak (see Table~\ref{tab:data}) as the characteristic ejecta velocity---the amount of $B$-band flux suppression needed to explain the observed SED at 0.5 days is (3.2 $\pm$ 0.6) $\times$ 10$^{-5}$~ergs~s$^{-1}$~cm$^{-2}$~Hz$^{-1}$ for a 13000 $\pm$ 5000~K blackbody source, corresponding to an isotropic $B$-band energy deficit of (2.5 $\pm$ 0.5) $\times$ 10$^{25}$ ergs~s$^{-1}$~Hz$^{-1}$.

Such a selective suppression of flux can only be from line absorption. Fe-peak and associated nucleosynthetic elements, including \ni56, \fe52, \chr48, \ti44, and their decay products, are the elements expected in Type Ia SNe that can significantly suppress flux in the vicinity of $\lesssim$ 5000~\AA.
While H and He lines can blanket $\lesssim$ 4000~\AA\cite{Mihalas1967apj, Mihalas1968apj}, they are not generally expected in \tase\ and most of the $B$-band flux would not be significantly affected.
We find that a distribution of Fe-peak elements near the ejecta surface can reproduce the $B$-band suppression, offering the best explanation for the infant-phase $B$-band plateau and red $B-V$ color of \uname.
In Figure~\ref{fig:specfig}, the observed SED at 0.5 days ($\sim$ 0.9 days post-explosion) is compared to the modelled spectrum of a \tas\ with an excess of Fe-peak elements in the outer 1\% of the SN-ejected mass at a nearly identical phase (1.05 days post-explosion), 
obtained by fitting the early light curves of \uname\ with a grid of He-shell DDet simulations (see below).
Indeed, the model predicts a $\sim$ 5000~\AA\ spectroscopic break with significant flux suppression bluewards of the break caused by Fe-line blanketing.
Estimating the amount of suppression by performing photometry on the modelled spectrum using the KMTNet filters (Figure~\ref{fig:specfig}, top panel), we obtain an isotropic $B$-band energy deficit of 2.1 $\times$ 10$^{25}$ ergs~s$^{-1}$~Hz$^{-1}$ relative to the $Vi$-band continuum of the model, consistent with what is required to explain the infant-phase SED of \uname.

{\noindent \bf Distribution of Fe-peak Elements in the Extreme Outer Ejecta}

To understand the origin of Fe-peak elements near the ejecta surface in \uname, we investigate if the elements
are part of the centrally-concentrated and monotonic main distribution of \ni56\ in the ejecta or if they constitute a distinct radial over-density in a shell (or clump, depending on the 3D distribution).
We find that the early light curves of the SN accommodate the presence of a radioactive shell/clump as follows.
Figure~\ref{fig:early} compares the observed early light curves to those predicted by blackbody models powered by four \ni56\ distributions: a logistic function\cite{Piro&Nakar2014apj} peaked at the ejecta core fitted to the SN light curves during 0--10 days (Supplementary Section 5), representing the main \ni56\ distribution, and the same function with 
total \ni56\ mass in the outer 0.31\% of the SN-ejected mass increased to (0.9, 1.8, and 3.6) $\times$~10$^{-4}$~\msol.
During 1--7 days, the logistic model follows similar rising behavior as the SN luminosity, confirming that the \ni56\ distribution is mainly centrally-concentrated and monotonic under the ejecta surface.
In the infant phase between 0 and 1 days, however, 
the pure logistic model under-predicts the observed luminosity in all three $BVi$ bands.
Excess radioactive materials is one possible explanation for the luminosity difference, as shown by the models with increased \ni56\ fractions.

The stratification of Fe-peak elements in the ejecta also affects the $B-V$ color of the SN\cite{Magee2020aa, Polin2019apj} as their absorption features (Figure~\ref{fig:specfig}) evolve over time.
Figure~\ref{fig:colorobs} shows that the $B-V$ color evolution of \uname\ is best-explained if the ejecta possesses an over-density of surface Fe-peak elements in addition to a centrally-concentrated and monotonic \ni56\ distribution.
The observed color curve is compared to those predicted by radiative transfer calculations of two different ejecta profiles: (1) a purely logistic \ni56\ distribution\cite{Magee2020aa} fitted to the light curves of \uname\ during 1--10 days (Supplementary Section 7.2) and (2) a logistic-like \ni56\ distribution with a shell of Fe-peak elements in the outer 1\% of the ejected mass from the best-fit He-shell DDet simulation (see below).
The model with an Fe-peak shell provides a much better fit to the observed color evolution from the infant phase to later epochs than the one with a logistic distribution alone, though both models are consistent with the observed color evolution after $\sim$ 2 days.

{\noindent \bf Nature of \uname\ and Implications for \tas\ Origins}

The presence of an Fe-peak shell/clump near the ejecta surface, as found in \uname, has critical implications for the normal Type Ia explosion mechanism.
Simulations of \tase\ have shown that over-densities of Fe-peak elements in the extreme outer ejecta can result from either mixing during a sub-sonic explosion process or surface nuclear burning during the explosion.
The former is predicted by some Chandrasekhar-mass explosion models, where a WD initially deflagrates sub-sonically before transitioning into a detonation\cite{Reinecke2002aa}.
When the explosion is spherically symmetric, 
the deflagration phase is usually limited to the inner ejecta while a detonation is required to traverse the low-density outer ejecta\cite{Nomoto2017hsn}, a process that is not expected to produce surface Fe-peak clumps\cite{Maeda2010apj}. 
However, some simulations have found that asymmetric explosions resulting from off-center deflagrations can produce surface Fe-peak clumps that are visible from a limited set of favourable viewing angles\cite{Maeda2010apj, Seitenzahl2013mnras}, which could be one possible explanation for the distribution of Fe-peak elements in SN 2018aoz.
Separately, a gravitationally confined detonation, where the off-center deflagration plume rises buoyantly to the progenitor surface, is another scenario that produces a spherical shell of Fe-peak elements near the ejecta surface, though recent simulations have shown that this produces peculiar explosions that are incompatible with normal \tase\cite{Seitenzahl2016aa}.

Sub-Chandrasekhar-mass He-shell DDets offer another way to produce Fe-peak elements in the outer ejecta.
In the case of the spectroscopically peculiar Type Ia SN MUSSES1604D\cite{Jiang2017nat}, the detonation of a
0.054$\,$\msol\ He-shell
was invoked to explain its redward color evolution during 1--2 days post-explosion (Figure~\ref{fig:colorobs}). 
To investigate if this scenario can explain a redward color evolution with a much shorter timescale and larger amplitude, as observed in \uname, we perform a grid of He-shell DDet \tas\ simulations with a range of WD and He-shell masses (Supplementary Section 7.1), fitting the observed light curves during 0--8 days.
We obtain the best-fit with a small He-shell mass of $0.01\,$\msol\ on a $1.05\,$\msol\ WD, which also provides the best match to the near-peak $BVi$ light curves and spectroscopic features.
As shown above, this model can indeed explain the $B$-band suppression (Figure~\ref{fig:specfig}, magenta spectrum) and $B-V$ color evolution (Figure~\ref{fig:colorobs}, magenta curve) of \uname.
If He-shell DDet is the origin of surface Fe-peak elements in \uname, it
would imply that detonations of He-shells as thin as $\sim$ 0.01\,\msol\ can successfully initiate normal \tase, consistent with the predictions of our simulations and other recent simulations\cite{Shen&Moore2014apj, Townsley2019apj}.
In our best-fit He-shell DDet model for \uname, 12.9\% of the outer 1\% of the ejecta by mass is composed of \ni56, \fe52, and \chr48.
Although this is comparable to the amount of radioactive material needed to reproduce the infant-phase excess emission in \uname\ (see section ``Distribution of Fe-peak Elements in the Extreme Outer Ejecta''), the He-shell DDet model under-predicts the emission $\lesssim$ 0.5 days (Supplementary Section 7.1), indicating that either the current He-shell DDet models do not fully capture radioactive heating at early times or another source of emission is also required (e.g., ejecta collision with a companion\cite{Kasen2006apj}).
Future modelling is necessary to ascertain how the inclusion of more detailed effects would influence the inferred mass of surface Fe-peak elements in \uname.

No other \tas\ has been detected in multiple bands at early enough epochs and sufficient depth to identify the remarkable $B$-band plateau and extreme redward $B-V$ color evolution observed during the first $\sim$ 1 day post-explosion in \uname. 
Thus, such features could be present in many normal \tase, but likely not all of them.
Observationally, based on the color evolution in later epochs (2--5 days; see Figure~\ref{fig:colorobs}), \uname\ appears to belong to the ``early-red'' population of Type Ia\cite{Stritzinger2018apj} comprising just over half of the normal events, although whether the reported ``early-red/blue'' dichotomy of \tase\ truly represents separate populations is still uncertain\cite{Stritzinger2018apj, Han2020apj, Bulla2020apj}.
Theoretically, we note that two of the three \tas\ explosion mechanisms described above that can accommodate the presence of surface Fe-peak elements can only account for some fractions of the entire normal Type Ia population in current models:
He-shell DDets leave almost no unburnt carbon, incompatible with $>$ 40\% of \tase\cite{Maguire2014mnras}, and the strong asymmetry of the gravitationally confined detonation scenario only leaves a small (if existent) set of compatible viewing angles for normal events\cite{Seitenzahl2016aa}.
\uname\ provides the first evidence that surface Fe-peak elements are required in at least a fraction of normal \tas\ explosions.

\clearpage

\begin{methods}

\subsection{Discovery and Photometric Observations of \uname.}

\uname\ was identified by both the KMTNet Supernova Program\cite{Moon2016spie, Afsariardchi2019apj, Moon2021apj} (KSP) and Distance Less Than 40 Mpc Survey\cite{Tartaglia2018} (DLT40). KSP uses the three 1.6m telescopes of the KMTNet in Chile, South Africa, and Australia in order to conduct a survey optimized for detecting and continuously monitoring infant SNe in multiple colors. Each telescope of the network is equipped with an identical wide-field CCD camera with 4 square degree field-of-view and multiple filters.
Between 2018 February and July, we conducted high-cadence monitoring of a 2\degr$\times$ 2\degr\ field containing the nearby elliptical galaxy NGC~3923. We obtained $\sim$ 700 images of the field with 60-s exposure times at a mean cadence of 4.7 hours in each of the $BVI$ bands. (Note that the $I$-band observations are calibrated to AAVSO $i$-band magnitudes as described in section ``Photometric Calibration'' below.) The typical limiting magnitude for a point source in these images is 21$-$22 mag at a signal-to-noise ratio (S/N) of 3. DLT40 utilizes the 0.4m PROMPT5 telescope\cite{Reichart2005} at Cerro Tololo Observatory in Chile to conduct a daily cadence survey of nearby galaxies, also optimized for detecting SNe at an early phase. The survey observations are unfiltered with a typical single-epoch depth of 19$-$20 mag. 

The earliest detections (S/N $>$ 3) of \uname\ were made by KSP at 00h54m on 29 March 2018 UT (MJD 58206.0378) in the $B$ and $V$ bands obtained with the Chilean KMTNet telescope at the coordinate (RA, decl.) = ($\rm 11^h51^m01^s.80$, $-28\degr44\arcmin38\farcs.5$) (J2000), 3.72\arcmin away from the center of \ngc in the northern direction. 
Supplementary Figure~1 (column 2) shows the first $BVI$-band images, taken within 4 minutes of each other, obtaining S/N of 3.5, 4.3, and 1.9, respectively, at the source position. 
The apparent magnitudes of the source in the images were measured to be 21.57$\,\pm\,$0.44, 21.26$\,\pm\,$0.25, and 21.97$\,\pm\,$0.58 mag, respectively, where the error includes contributions from background noise at the source position, photometric calibration, and $B$-band S-correction (see below).
The source was not detected in $B$-, $V$-, or $I$-band images obtained 1.1 days before the first detection. DLT40 detected the source 1.1 days later
and made the discovery report at 07h25m on 2 April 2018 UT (MJD 58210.4158)\cite{Valenti2018, Brown2018at}. 

In addition to observations made by KSP and DLT40, we conducted optical photometric observations of \uname\ using 1m telescopes of the Las Cumbres Observatory (LCO) network of robotic telescopes in conjunction with the Global SN Project between 4.4 and 64 days since first light.\cite{Brown2013pasp} These observations are supplemented by UV observations acquired by the Neil Gehrels Swift Observatory\cite{Gehrels2004} Ultra-Violet Optical Telescope\cite{Roming2005ssr} (UVOT), and NIR observations from the ANDICAM\cite{DePoy2003spie} instrument on the SMARTS 1.3m telescope at Cerro Tololo Observatory. The Swift-UVOT and ANDICAM observations were carried out during the periods of 5.5$-$59 and 9.1$-$110 days since first light, respectively.

\subsection{Photometric Calibration.}

Point-spread function (PSF) photometry of \uname\ on KSP images was performed using the SuperNova Analysis Package (SNAP; Code Availability), a custom python-based pipeline for SN photometry and analysis.
A local PSF was obtained by fitting a Moffat function\cite{Moffat1969aap, Trujillo2001mnras} to nearby reference stars and simultaneously fitting sky background emission with a first-order polynomial function. 
The fluxes of \uname\ were obtained by fitting the local PSF near the source location, and the detection S/N is equal to the best-fit flux divided by its 1-$\sigma$ uncertainty.
Photometric flux calibration was performed against 6--9 standard reference stars within 10\arcmin of the source from the AAVSO Photometric All-Sky Survey\cite{Henden2018aas} database whose apparent magnitudes are in the range of 15--16 mag; the observations in the $BVI$ KMTNet filters were calibrated against reference stars in the nearest AAVSO filters (Johnson~$BV$, and Sloan $i'$; or $BVi$).

The KSP instrumental magnitudes for the AAVSO reference stars were transformed to standard $BVi$ magnitudes using the equations from Park et al. (2017)\cite{Park2017apj}.
However, the spectra of SNe, particularly after the peak, are significantly different from the AAVSO standard stars used to derive the transformation equations. In order to account for this, we applied spectrophotometric (S)--corrections, which are magnitude corrections between instrument and standard filters derived by performing synthetic photometry on spectra obtained at the same epoch\cite{Stritzinger2002aj}. 
We applied interpolated S--corrections to observations following the first spectrum taken at UT 09h22m on April 2 (MJD 58210.39), 4.4 days since first light.
At epochs observed before 4.4 days, we applied the Park et al. (2017) transformation equations to the source in lieu of S--corrections under the assumption that the spectrum of \uname\ in the early phase is dominated by continuum emission, which is supported by the small magnitude of S--corrections before the peak.

We searched for faint background sources around the location of the SN by using $\gtrsim$ 300 individual pre-SN images in each of the $BVI$ bands to create deep stacked images shown in Supplementary Figure~2 (right panels).
As seen in the figure, we identified a background source $\sim$ 0.8\farcs\ north-east of the SN position with apparent $BVi$-band magnitudes of 24.90$\,\pm\,$0.27, 24.02$\,\pm\,$0.20, and 22.39$\,\pm\,$0.08 mag, respectively. 
(See section ``Host, Distance, and Extinction'' below for the discussion of the nature of this source).
In the $B$ and $V$ bands, the brightness of the source is significantly below the 1-$\sigma$ detection limits of the earliest images within 0.5 days---22.9 and 22.7 mag for $B$- and $V$-band detection images, respectively---and much smaller than that of the SN at any epoch ($<$ 21.8 mag for $B$ band and $<$ 21.3 mag for $V$ band). 
Therefore, we ignore the background source in the $B$- and $V$-band photometry as it is incapable of meaningfully affecting the PSF photometry of \uname. 
For the $i$-band photometry, we remove the background source when the SN brightness is $\lesssim$ 100 times that of the background source (which is only for epochs before $\sim$ 0.5 days) by measuring the SN flux using a Kron aperture containing both sources and subtracting the known flux of the background source.

Unfiltered observations obtained with the DLT40 survey were template subtracted and then calibrated to APASS $r$-band observations of field stars.
Additional optical observations with the LCO 1-m telescopes were carrried out in the $U$, $B$, $V$, $g$, $r$, and $i$ bands. Photometric reduction for the LCO images was accomplished using \texttt{lcogtsnpipe}\cite{Valenti2016}, a PyRAF-based pipeline.
Zeropoints for images in Landolt filters were calculated from Landolt standard field images taken by the telescopes on the same nights as the observations. Zeropoints for images in Sloan filters were calculated using APASS stars in the same field-of-view as \uname.

Observations with the Neil Gehrels Swift Observatory were triggered under the Swift Guest Investigator key project Swift Response to Nearby Supernovae.  
Swift observations began
at 13h57m on 3 April 2018 UT (MJD 58211.5811),
and used all six medium band filters--$UVW2$, $UVM2$, $UVW1$, $U$, $B$, and $V$.  Coaddition of same-day images and photometry were done using the pipeline of the Swift Optical Ultraviolet Supernova Archive\cite{Brown2014a} (SOUSA) and the photometry is calibrated to the UVOT Vega system\cite{Breeveld2011}.
The Swift/UVOT photometric reduction includes the subtraction of the host galaxy count rates using images obtained August 2019.  The UVOT photometry uses the September 2020 update to the time-dependent sensitivity correction\cite{breeveld2020}.

Seventeen epochs of $J$-, $H$-, and/or $K$-band observations of \uname\ were obtained with ANDICAM mounted to the 1.3m telescope at the Cerro Tololo Inter-American Observatory (CTIO) and operated by the SMARTS Consortium under programme NOAO-18A-0047. For each epoch, a set of 10 individual 20s images were taken per band, with a hexagonal 7-position dithering of around 20 arcsec.
The dithered frames were stacked to produce the final co-added images for each epoch. PSF photometry was performed on the SN and nearby field stars using standard \texttt{daophot} packages within the Image Reduction and Analysis Facility (IRAF) software and absolute calibration was performed using Two Micron All Sky Survey\cite{Skrutskie2006} (2MASS) $JHK$ magnitudes. 

The light curves of the SN from all the photometric observations are shown in Supplementary Figure~3 and provided in Supplementary Table~1.
The photometry will also be publicly available on the Open SN Catalog\cite{Guillochon2017apj}.

\subsection{Robustness of Infant-phase KSP Photometry.}

Supplementary Figure~1 presents the $B$- , $V$-, and $I$-band stamp images (rows 2--4) showing the detection of the source at the epoch of the first detection (column 2), three following epochs during its infant phase ($\lesssim$ 0.5 days since first light; columns 3--5), and one epoch immediately before and after the infant phase (columns 1 and 6, respectively), as well as a table (row 1) showing the time and moon position at each observation. 
The moon was nearly full and at 25--45\degr elevation during the epochs, contributing to the background noise in the images. 
We find $B$-band 3-$\sigma$ detection limits of 20.8--21.9 mag in single-epoch images from 0--0.5 days, consistent with typical limiting magnitudes for the KMTNet\cite{Jeong2016jkas} and significantly deeper than the $B$-band plateau at multiple epochs in the phase, despite the moon brightness.
Note that these limiting magnitudes are estimated in a conservative way from background pixel variations, which gives limits that are slightly shallower than the depths detectable using PSF photometry in our data.
Supplementary Figure~1 provides the measured S/N at the SN position in the bottom-right corner of each stamp image, showing the robust (S/N $>$ 5) detection of the SN in all three bands at 0.11, 0.18, and 0.51 days during its $B$-band plateau and redward $B-V$ color evolution.

We confirm the detection of the infant-phase $B$-band suppression as follows. 
Supplementary Figure~1 (column 5) shows stamp images of the SN at 0.51 days, when its $B$-band flux was lowest compared to the $V$ and $i$ bands and its $B-V$ color reached maximum.
The images were obtained by binning 2--3 adjacent images within 1.5 hours of each other for higher S/N.
Each of the $B$-, $V$-, and $I$-band images robustly detected the SN with S/N $>$ 7.
In these images, we measure $BVi$ magnitudes of the SN to be 21.78 $\pm$ 0.21, 20.02 $\pm$ 0.04, and 20.33 $\pm$ 0.08 mag, respectively, corresponding to a $B$-band flux that is 3.7 $\pm$ 0.7 times lower than the average of the $V$ and $i$ bands, after applying extinction correction.
The detection of the $B$-band suppression is also confirmed with synchronous $BVi$ fluxes of the SN at 0.5 days (Figure~\ref{fig:specfig}, yellow stars) obtained via Gaussian process light curve interpolation (see below), which show a $B$-band flux that is 4.4 $\pm$ 1.2 times lower than the average of the $V$ and $I$ bands.
We note that it is impossible to reconcile this difference with a modification to the flux of the background source located $\sim$ 0.8\farcs\ from the position of \uname. 
The background source is 18 times dimmer than the SN at this epoch in $B$-band and incapable of affecting the SN brightness significantly even if it vanished. 
Conversely, the background source brightness would have needed to increase by factors of 31 and 5.5 in the $V$ and $I$ bands, respectively, relative to its brightness in our deep stacked images---without a corresponding increase in the $B$ band---in order to lower the inferred $V$- and $i$-band fluxes of the SN at this epoch to that of the $B$ band. This is not supported by the centroids of the $V$- and $I$-band SN detections at this epoch, which are aligned with the SN position ($<$ 0.04\farcs\ away). We therefore conclude that the measured suppression of the $B$-band flux relative to that of the $V$ and $I$ bands is robust.

\subsection{Spectroscopy of \uname.}

We obtained 25 low-resolution optical spectra of \uname\ spanning 4.4--136.3 days since first light
with a combination of the Gemini Multi-Object Spectrograph\cite{Hook2004} (GMOS) on the 8.1m Gemini-South telescope, the Wide-Field CCD (WFCCD) on the 2.5m du Pont telescope at Las Camapanas Observatory, and the FLOYDS spectrographs located on the 2m Faulkes telescopes of the Las Cumbres Observatory in Siding Spring and Haleakala\cite{Brown2013pasp}. Two moderate-resolution spectra were obtained near maximum light in the region around Na~I~D with the Blue Channel Spectrograph\cite{Schmidt1989} on the MMT Telescope. The spectroscopic observations are summarized in Supplementary Table~2.

Spectra from the du Pont and MMT Telescopes were reduced using standard tasks within IRAF. Bias and flat-field corrections were performed on the two-dimensional frames, one-dimensional spectra were extracted, and wavelength calibration was performed using calibration lamps taken immediately after target exposures. Flux calibration and telluric corrections were peformed with a set of custom IDL scripts\cite{Matheson2008,Blondin2012aj} using spectrophotometric standards observed on the same night. GMOS spectra were reduced in a similar manner, but using the custom \texttt{gmos} suite of IRAF tasks. Initial flux calibration for GMOS spectra was performed using the IRAF tasks \texttt{standard} and \texttt{calibrate}, and final scaling was performed based on matching to observed photometry. 

A selection of our best low resolution spectra are presented in Supplementary Figure~4 (left panel), as well as the two moderate resolution spectra.
(The full set of spectra obtained will be available on the Open SN Catalog\cite{Guillochon2017apj} and WISeREP\cite{Yaron2012pasp}).
As seen in the figure, the spectra of \uname\ are dominated by high-velocity absorption features of Ca II, Fe II/III, S II, and Si II at early times. 
In particular, the presence of silicon features and the absence of hydrogen features confirm its \tas\ nature.
Supplementary Figure~4 (top-right panel) shows that the spectrum of the SN near peak is nearly identical to normal \tase\ 1994D\cite{Meikle1996mnras} and
2002dj\cite{Pignata2008mnras}.

\subsection{Host, Distance, and Extinction.}

The host galaxy of \uname\ is NGC~3923, an elliptical galaxy (Supplementary Figure~2, left panel), as confirmed below by their matching redshifts and distance moduli.
Using SNID\cite{Blondin2007apj}, we find a best-fit spectral template to the maximum light spectra of \uname\ with a spectroscopic redshift of $z$ = 0.0058 $\pm$ 0.0007,
which matches the known redshift of the galaxy\cite{Norris2011mnras}, $z$ = 0.00580 $\pm$ 0.00003.
Using NED\cite{Steer2017aj}, we determine a distance modulus (DM) of 31.75 $\pm$ 0.15 mag to the SN based on Hubble flow in the cosmology of Riess et al. (2016)\cite{Riess2016apj} with corrections for peculiar velocities due to the Virgo Supercluster, Great Attractor, and Shapley Supercluster\cite{Mould2000apj}.
This value is consistent with redshift independent distances to NGC~3923 based on surface brightness fluctuations (DM = 31.64 $\pm$ 0.14 mag)\cite{Tully2013aj}, and to \uname\ based on \tas\ template fitting (DM = 31.75 $\pm$ 0.08 mag; see next section).
The luminosity distance to the SN based on the \tas\ template DM is 22.4 $\pm$ 0.8~Mpc.
The location of the SN $\sim$ 3.72\arcmin north of the galaxy center (Supplementary Figure~2), corresponding to a projected distance of 24.0~$\pm$~0.9~kpc, places it in the stellar halo of the galaxy.

\uname\ suffers from relatively little Galactic extinction since it is located near the Galactic south pole with $E(B-V)$ = 0.093 $\pm$ 0.001 mag in its direction according to the extinction model of Schlafly \& Finkbeiner (S\&F 2011)\cite{Schlafly&Finkbeiner2011apj}, corresponding to extinction correction of 0.40, 0.29, and 0.18 mag for the $B$, $V$, and $i$ bands, respectively, for an $R_V$ = 3.1 Fitzpatrick (1999) reddening law\cite{Fitzpatrick1999pasp}.
For the spectra and spectral energy distribution of the SN (Supplementary Information), we adopted Galactic extinction corrections based on the updated reddening law of Fitzpatrick \& Massa (F\&M 2007)\cite{Fitzpatrick&Massa2007apj}, which is slightly better in the UV regime.
Given the halo environment of \uname, little extragalactic extinction from NGC~3923 is expected.
This is confirmed by the lack of observed Na I doublet (Na I D; 5890 and 5896~\AA) absorption features at the redshift of the SN in two moderate resolution MMT spectra (Supplementary Figure~2, inset).
We confirm the Galactic extinction to the source by fitting the equivalent widths of the Na~I~D in the observer frame using a Voigt doublet profile in each of the two MMT spectra\cite{Poznanski2012mnras}.
The mean of the measurements is $E(B-V)$ = 0.07 $\pm$ 0.01 mag, consistent with the value from S\&F 2011. 
The non-detection of Na~I~D in the rest-frame of the SN in these spectra constrains the host galaxy extinction to be significantly below this value.
By scaling the measured Voigt profiles of the Galactic Na I D to the MMT spectra at $z$ = 0.0058, we estimate a 3-$\sigma$ upper limit on the equivalent width of Na I D to be 0.11~\AA, corresponding to $E(B-V) <$ 0.02 for the host galaxy, assuming a Milky-Way-like correlation between Na I D and dust extinction\cite{Poznanski2012mnras}.

The nature of the background source located $\sim$ 0.8\farcs\ north-east and underlying the position of \uname\ in KSP images (Supplementary Figure~2) is uncertain, though we consider several possibilities below.
It can be a source at the redshift of the SN, in which case the low $V$-band luminosity ($-$7.7 mag) and red $V-I$ color ($-$1.6 mag) of the source are consistent with it being a dwarf galaxy or globular cluster in NGC~3923\cite{Misgeld&Hilker2011mnras, Gavazzi2010aa, Harris1991apjs}.
The projected distance of $\sim$ 86~pc between the SN and the source in the rest frame opens the possibility for association between the two, though the source apparently does not provide a dusty host environment, as evidenced by the lack of Na~I~D lines in the rest frame (see above).
Alternatively, the source can be a spatially coincident foreground star or background galaxy.
Additional observations with higher spatial resolution are required to determine whether the source is separated from the SN or truly underlying.

\subsection{Light Curve Parameters.}

The light curves of \uname\ (Supplementary Figure~3) display two post-peak decay timescales, consistent with SNe powered by \ni56\ and \co56\ radioactive decay. As typically found in normal \tase, each of the NIR $iJHK$-band light curves reach a primary peak before the $B$ and $V$ bands, followed by a secondary peak associated with the recombination of iron group elements in the ejecta\cite{Kasen2006apj}. 
The light curves of normal \tase\ in the photospheric phase from $-$10 to 15 days since peak are also known to form a family of functions parameterized by the Phillips parameter, \dm15, which measures the post-peak decline rate in the $B$-band light curve\cite{Phillips1999aj}.
To effectively include some peculiar \tase, such as the rapidly evolving 91bg-like subtype, 
the color stretch parameter, \sbv, has been used as an alternative to \dm15, defined by $s_{BV} = t_{BV}/($30 days$)$, where $t_{BV}$ is the time between $B$-band peak to the maximum post-peak $B-V$ color\cite{Burns2014apj}.
We used SNooPy\cite{Burns2011aj} to fit the $BgVri$-band light curves of \uname\ with a normal \tas\ template based on Hsiao et al. (2007)\cite{Hsiao2007apj}, where the fitted parameters are stretch, time of $B$-band peak (\tp), and DM.
In the fitting process, we applied $K$--correction between the observer and rest-frame filter response functions for each filter (see below) and correct for Galactic extinction.
The best-fit template (see Supplementary Figure~3) provides a good comparison to our observed light curves with \tp\ and DM of $58222.27\pm0.02$ (MJD) and $31.75\pm0.08$ mag, respectively.
The stretch parameter of the best-fit template is 0.87 $\pm$ 0.03.

We derive rest-frame magnitudes for \uname\ by adopting the DM and $K$--corrections from the best-fit template and Galactic extinction correction.
The $K$--corrections are insubstantial ($<$ 0.02 mag) given the relatively small redshift of the SN. 
With rest-frame light curves for the SN, we used polynomial fitting to measure the peak absolute magnitudes of each $BVi$-band light curve to be $-19.319\pm 0.009$, $-19.226\pm0.009$, and $-18.614\pm0.012$ mags, respectively.
The corrected $BVi$-band light curves each attained their peaks at $-0.86$, $-0.97$, and $-3.98$ days (in the observer frame) prior to \tp\ obtained from template fitting above, respectively.
We also measured the Phillips parameter and color stretch parameter to be \dm15\ $= 1.117\pm0.015$ mag and \sbv\ $= 0.797\pm0.019$, respectively.
These parameters are close to the average of normal \tase\cite{Parrent2014apss, Burns2014apj}, while they are inconsistent with subluminous (91bg-like) and overluminous (91T-like) events, confirming the classification of \uname\ as a normal \tas.

\subsection{Data Availability}

The source data for Figure~\ref{fig:lc} during 0--1 days since first light is available in the online version of this paper, while the data for the entire single-epoch UV to NIR light curves are provided in the Supplementary Information. All photometric and spectroscopic data are also available on the Open Supernova Catalog\cite{Guillochon2017apj} and WISeREP3\cite{Yaron2012pasp}. 
The modelled light curves and spectra of our He-shell DDet simulations are available at https://github.com/niyuanqi/he-shell-ddet.

\subsection{Code Availability}

We performed light curve template fitting in the post-infant phase using SNooPy, available at https://csp.obs.carnegiescience.edu/data/snpy/snpy.
In our He-shell DDet models, hydrodynamics and nucleosynthesis simulations were conducted using Castro\cite{Almgren2010apj, Zingale2018jphc}, and the radiative transfer calculations were conducted using Sedona\cite{Kasen2006bapj}.
The code used to measure the KMTNet light curves of \uname, construct the bolometric light curves, and generate the analytic \ni56-powered light curve models are available at https://github.com/niyuanqi/SNAP.
IRAF is distributed by the National Optical Astronomy Observatory, which is operated by the Association of Universities for Research in Astronomy, Inc. under cooperative agreement with the National Science Foundation.

\end{methods}

\begin{addendum}
 \item This research has made use of the KMTNet system operated by the Korea Astronomy and Space Science Institute (KASI) and the data were obtained at three host sites of CTIO in Chile, SAAO in South Africa, and SSO in Australia. 
 The Gemini South observations were obtained under the K-GMT Science Program (PID: GS-2018A-Q-117 and GS-2018B-Q-121) of KASI. 
 The Swift observations were triggered through the Swift GI program 80NSSC19K0316. SOUSA is supported by NASA's Astrophysics Data Analysis Program through grant NNX13AF35G. 
 Some of the data presented herein were obtained at the W. M. Keck Observatory, which is operated as a scientific partnership among the California Institute of Technology, the University of California and the National Aeronautics and Space Administration. The Observatory was made possible by the generous financial support of the W. M. Keck Foundation. The Computational HEP program in The Department of Energy's Science Office of High Energy Physics provided simulation resources through Grant \#KA2401022. This research used resources of the National Energy Research Scientific Computing Center, a U.S. Department of Energy Office of Science User Facility operated under Contract No. DE-AC02-05CH11231.
 D.-S.M., M.R.D., and C.D.M. are supported by Discovery Grants from the Natural Sciences and Engineering Research Council of Canada.
 D.-S.M. was supported in part by a Leading Edge Fund from the Canadian Foundation for Innovation (project No. 30951).
 M.R.D. was supported in part by the Canada Research Chairs Program, the Canadian Institute for Advanced Research (CIFAR), and the Dunlap Institute at the University of Toronto.
 D.J.S. acknowledges support by NSF grants AST-1821987, 1821967, 1908972 and from the Heising-Simons Foundation under grant \#2020-1864. 
 S.G.-G. acknowledges support by FCT under Project CRISP PTDC/FIS-AST-31546 and Project UIDB/00099/2020.
 H.S.P. was supported in part by the National Research Foundation of Korea (NRF) grant funded by the Korean government (MSIT, Ministry of Science and ICT; No. NRF-2019R1F1A1058228).
 P.J.B. acknowledges support from the Swift GI program 80NSSC19K0316.
 S.V., Y.D., and K.A.B. acknowledge support by NSF grants AST-1813176  and AST-2008108.
 C.M. acknowledges support by NSF grant AST-1313484. 
 I.A. is a CIFAR Azrieli Global Scholar in the Gravity and the Extreme Universe Program and acknowledges support from that program, from the Israel Science Foundation (grant numbers 2108/18 and 2752/19), from the United States - Israel Binational Science Foundation (BSF), and from the Israeli Council for Higher Education Alon Fellowship.
 R.L.B. acknowledges support by NASA through Hubble Fellowship grant \#51386.01 awarded by the Space Telescope Science Institute, which is operated by the Association of  Universities for Research in Astronomy, Inc., for NASA, under contract NAS 5-26555.
 A.G.-Y. acknowledges support from the EU via ERC grant No. 725161, the ISF GW excellence center, an IMOS space infrastructure grant and BSF/Transformative and GIF grants, as well as The Benoziyo Endowment Fund for the Advancement of Science, the Deloro Institute for Advanced Research in Space and Optics, The Veronika A. Rabl Physics Discretionary Fund, Paul and Tina Gardner, the Yeda-Sela Center for Basic Research and the WIS-CIT joint research grant. A. G.-Y. is the recipient of the Helen and Martin Kimmel Award for Innovative Investigation. 
 L.G. acknowledges financial support from the Spanish Ministerio de Ciencia e Innovaci\'on (MCIN), the Agencia Estatal de Investigaci\'on (AEI) 10.13039/501100011033, and the European Social Fund (ESF) "Investing in your future" under the 2019 Ram\'on y Cajal program RYC2019-027683-I and the PID2020-115253GA-I00 HOSTFLOWS project, and from Centro Superior de Investigaciones Cient\'ificas (CSIC) under the PIE project 20215AT016.
 G.P. acknowledges support from the Millennium Science Initiative through grant IC120009.
 J.A. is supported by the Stavros Niarchos Foundation (SNF) and the Hellenic Foundation for Research and Innovation (H.F.R.I.) under the 2nd Call of ``Science and Society'' Action Always strive for excellence -- ``Theodoros Papazoglou’' (Project Number: 01431)
 
 \item[Author Contributions] 
 Y.Q.N. conducted most of the analyses under the supervision of D.-S.M. and M.R.D..
 D.-S.M. is the principal investigator of the KMTNet Supernova Program (KSP) that detected the infant-phase features of \uname\ and wrote the KSP pipeline.
 M.R.D. led collaboration between the KSP and other partners. 
 Y.Q.N. D.-S.M. and M.R.D. co-drafted the manuscript.
 A.P. conducted the He-shell DDet simulations under the supervision of P.N..
 D.-S.M., M.R.D., Y.Q.N., N.A., S.G.-G., S.C.K., Y.L., H.S.P., J.A., A.G.-Y., S.B.C., G.P. and S.D.R. are members of the KSP.
 N.A., D.-S.M., M.R.D., R.G.C. and C.D.M. are members of the Canadian Gemini South observing program for the KSP. 
 H.S.P., D.-S.M., S.C.K., and Y.L. are members of the Korean Gemini South observing program for the KSP.
 A.L.P. performed the shock breakout modelling. 
 P.J.B. led the Swift program for UV observations with help from S.B.C..
 L.G. and G.P. obtained the ANDICAM NIR observations.
 D.J.S. and S.V. co-led the DLT40 Program.
 J.H., D.E.R., V.K. and S.W. contributed to the operation of the DLT40 program.
 S.Y. built the machine learning implementation for the DLT40 survey. 
 K.A.B. Y.D., J.E.A. and N.S. are members of the DLT40 team who obtained the Keck and MMT spectra.
 S.-M.C. and Y.L. helped operate the KMTNet.
 D.A.H., C.M., I.A., J.B., D.H., and G.H. contributed to the LCO photometry and the FLOYDS spectroscopy. 
 M.R.D., R.L.B., T.W.-S.H., S.D.J., N.M., and J.R. contributed to the du Pont WFCCD and Magellan spectra.
 All of the authors contributed to discussion.

 \item[Competing Interests] The authors declare
no competing interests.
\end{addendum}

\newpage
\clearpage

\begin{table}
\centering
\medskip
\begin{tabular}{llr}
\hline\hline
& Parameter & Value\\
\hline
    1 & Sky coordinates (RA, decl.) (J2000)        & ($\rm 11^h51^m01^s.80, -28\degr44\arcmin38\farcs.48$) \\
    2 & Redshift ($z$)                             & 0.00580 $\pm$ 0.00003 \\
    3 & First detection: UT \& MJD                 & 00$h$ 54$m$ on March 29, 2018 \& MJD  58206.03779 \\
    4 & Observed peak magnitude                    & 12.81 ($B$), 12.80 ($V$), 13.26 ($i$) mag  \\ 
    5 & Observed peak epochs                                & MJD 58221.41 ($B$), 58221.30 ($V$), 58218.29 ($i$) \\
    6 & Early light curve power-law indices ($\alpha$)    &  2.24 $\pm$ 0.02 ($B$), 1.99 $\pm$ 0.02 ($V$), 2.26 $\pm$ 0.01 ($i$) \\  
    7 & Epoch of first light ($t_{\rm efl}$)               & MJD 58206.00 $\pm$ 0.02 \\
    8 & First detection from epoch of first light  & 1.0 $\pm$ 0.5 hours \\
    9 & Post-peak decline rate (\dm15)             & 1.117 $\pm$ 0.015 mag \\
    10 & Color stretch parameter (\sbv)             & 0.797 $\pm$  0.019  \\
    11 & Peak absolute magnitude                    & --19.32 ($B$), --19.23 ($V$), --18.61 ($i$) mag \\
    12 & Peak Si~II velocity                 &  (11.43 $\pm$ 0.12) $\times$ 10$^{3}$ km s$^{-1}$ \\
    13 & Explosion epoch ($t_{\rm exp}$)               & MJD 58205.6 $\pm$ 0.7    \\
    14 & Peak bolometric luminosity ($L_{\rm peak}$)  & 1.42 $\times$ 10$^{43}$ erg s$^{-1}$   \\
    15 & Bolometric peak epoch                      & --1.16 days \\
    16 & \ni56\ mass ($M_{\rm Ni}$)                 & 0.509 $\pm$ 0.006~\msol\ \\
    17 & $\tau_m$ parameter                        &
    9.51 $\pm$ 0.26~days \\
    18 & Ejecta mass ($M_{\rm ej}$)                & $\sim$ 0.8--1.0~\msol\  \\
    19 & Ejecta kinetic energy ($E_{\rm ej}$)       & $\sim$ (0.6--0.8) $\times$ 10$^{51}$ erg \\
\hline
\end{tabular}
\caption{\textbf{Properties of \uname.} The following properties are listed in the rows: (1) sky coordinates; (2) redshift of the host galaxy NGC~3923\cite{Norris2011mnras}; (3) first detection; (4) and (5) the observed peak apparent magnitudes and peak epochs of the $BVi$-band light curves; (6) and (7) the indices and epoch of first light from the power-law fit to the $BVi$-band early light curves (Supplementary Section 1); (8) the difference between the epochs of first light and first detection; (9), (10), and (11) \dm15, \sbv, and peak absolute magnitudes of the rest-frame $BVi$-band light curves (Methods); (12) peak Si~II velocity (Supplementary Section 2); (13) the explosion epoch from the power-law fit to the Si~II velocity evolution (Supplementary Section 2); (14) and (15) the peak luminosity and peak epoch of the bolometric light curve (Supplementary Section 3); (16), (17), (18), and (19) \ni56\ mass, $\tau_m$, ejecta mass, and ejecta kinetic energy from modelling the bolometric luminosity (Supplementary Section 4). All time/date values are given in the observer frame.}
\label{tab:data}
\end{table}

\newpage
\clearpage

\begin{figure}
\includegraphics{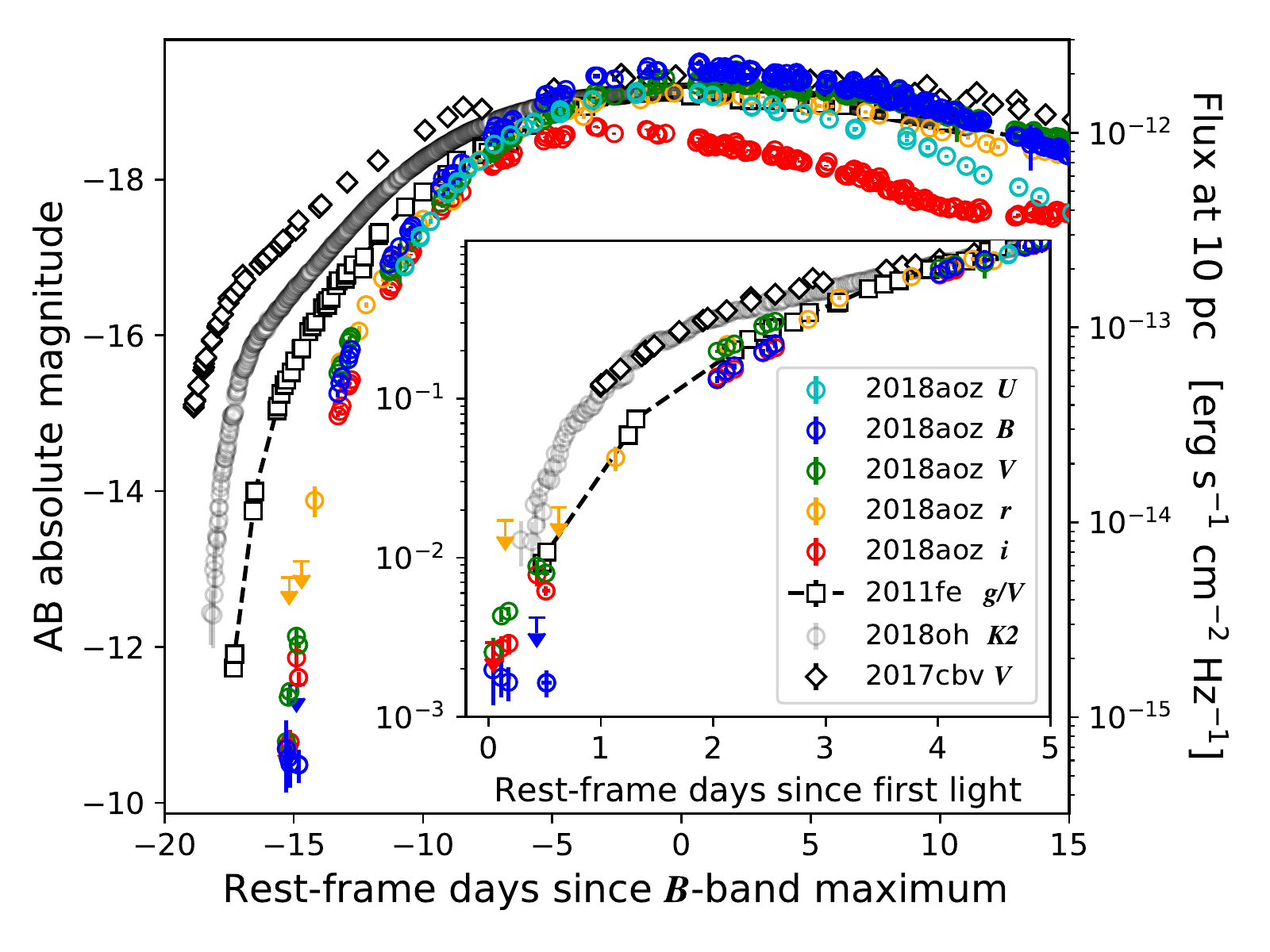}
\caption{\textbf{Subset of the light curves of \uname\ compared to three other early SNe.} 
The dereddened $UBVri$-band early light curves of \uname\ (colored circles) are compared to those of normal
\tase~2011fe\cite{Guillochon2017apj}, 
2018oh\cite{Dimitriadis2019apj}, and 2017cbv\cite{Hosseinzadeh2017apj} in their nearest-to-$V$ bands.
The errorbars represent the 1-$\sigma$ uncertainty level and the inverted arrows are 3-$\sigma$ detection limits.
The light curves of \uname\ show a faster pre-peak rise than the other three SNe, although they have similar post-peak decline rates.
The inset shows the early light curves of the SNe within 5 days of their epochs of first light, where excess emission has previously been identified in some \tase\cite{Dimitriadis2019apj, Hosseinzadeh2017apj, Stritzinger2018apj}, normalized to their luminosities at 5 days.
SN~2018oh and 2017cbv both show excess emission between 1 and 5 days, whereas SN~2011fe shows a typical power-law-like rise.
The light curves of \uname\ after $\sim$ 0.5 days are similar to SN~2011fe.
Between 0 and 0.5 days the light curves of \uname\ reveal the infant phase of \tas\ evolution.
In this phase, the $B$-band light curve stalls, showing a ``plateau'' feature, while the $Vi$-band ones rise rapidly at a rate of $\sim$ 2 mags/day.
}
\label{fig:lc}
\end{figure}

\begin{figure}
\includegraphics[width=1.0\textwidth,keepaspectratio]{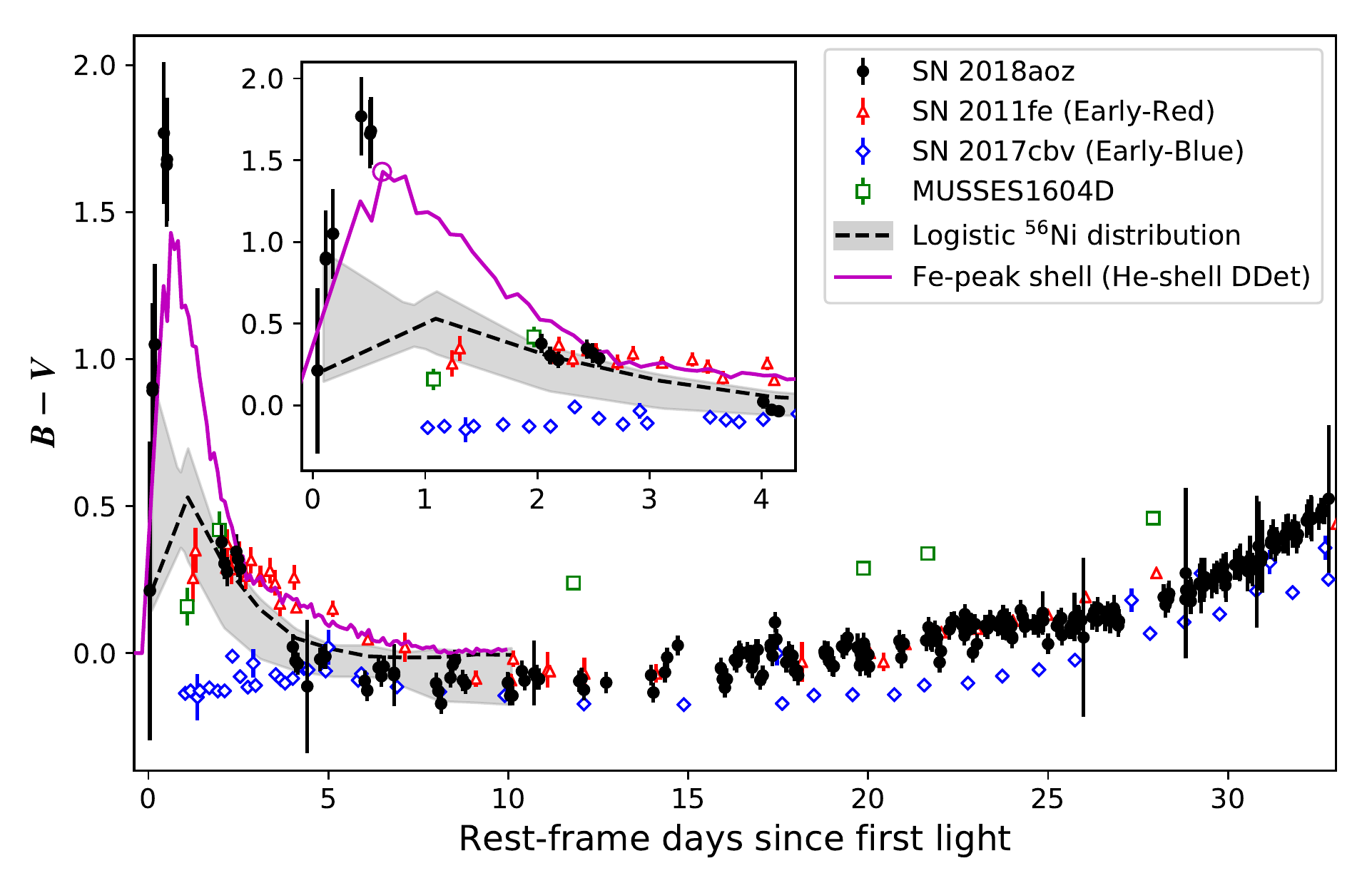}
\caption{\textbf{Comparison of the color evolution of \uname\ to other SNe and models.} 
The dereddened $B-V$ color of \uname\ (black filled circles) is compared to those of normal
\tase~2011fe\cite{Guillochon2017apj} 
and 2017cbv\cite{Hosseinzadeh2017apj} from the ``early-red'' and ``early-blue'' groups\cite{Stritzinger2018apj}, respectively, and MUSSES1604D\cite{Jiang2017nat}, a spectroscopically peculiar \tas\ with an early redward color evolution.
The errorbars represent the 1-$\sigma$ uncertainty level.
Also shown are color curves of fitted models from radiative transfer calculations of two different \ni56\ distributions: (1) a pure logistic \ni56\ distribution (black dashed curve with grey shaded error region; see Supplementary Section 7.2); and (2) a logistic-like \ni56\ distribution with Fe-peak elements in the outer 1\% of the SN-ejected mass (magenta solid curve) from a 0.01$\,$\msol\ He-shell $+$ 1.05$\,$\msol\ WD He-shell DDet simulation.
The $B-V$ color of \uname\ is obtained by linearly interpolating each of the $BV$-band light curves to the union of their epochs, while those of other SNe, 2011fe, 2017cbv, and MUSSES1604D, were obtained by binning each of their light curves with intervals of 0.05, 0.1, and 0.1 days, respectively, starting from their first observations.
The inset zooms in on the period of 0--4 days, where \uname\ shows an infant-phase redward color evolution within the first $\sim$ 0.5 days due to the $B$-band plateau.
The color evolution of \uname\ after $\sim$ 0.5 days is consistent with that of SN~2011fe and the logistic \ni56\ distribution. 
The magenta open circle shows the epoch of the simulated He-shell DDet spectrum in Figure~\ref{fig:specfig} (magenta spectrum).
}
\label{fig:colorobs}
\end{figure}

\begin{figure}
\includegraphics[width=1.0\textwidth,keepaspectratio]{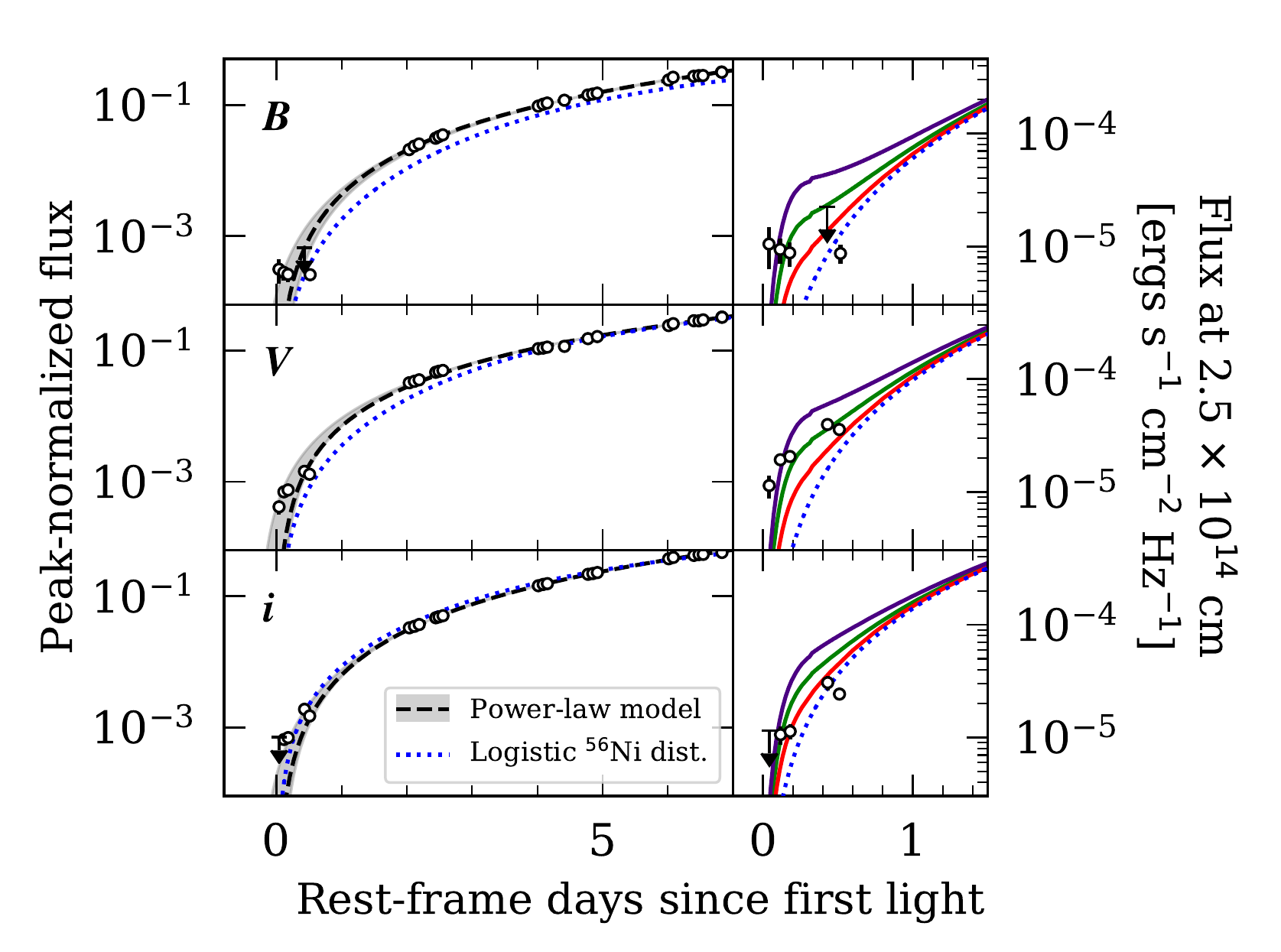}
\caption{\textbf{Early light curves of \uname\ compared to those of model predictions.} 
(Left) The rest-frame dereddened $BVi$-band (from top to bottom) light curves of \uname\ (circles) are compared to the best-fit power-law (black dashed curve) and the best-fit blackbody light curve model powered by a logistic \ni56\ distribution peaked at the ejecta core (blue dotted curve). 
The errorbars of the data points represent the 1-$\sigma$ uncertainty level and the inverted arrows are 3-$\sigma$ detection limits.
The upper and lower boundaries of the grey shaded region are the fitted power-laws when the infant-phase data during 0--0.5 days are entirely excluded and included, respectively.
(Right) The observed infant-phase fluxes of \uname\ before 0.5 days at its approximate photosphere radius are compared to the logistic \ni56\ model (blue dotted curve), along with modified logistic \ni56\ models (colored curves) where the
total \ni56\ mass in the outer 0.31\% of the SN-ejected mass has been increased to (0.9, 1.8, and 3.6) $\times$ 10$^{-4}$~\msol\ (red, green, and purple).
The increases in \ni56\ lead to excess emission before $\sim$ 1 day.
}
\label{fig:early}
\end{figure}

\begin{figure}
\includegraphics[width=1.0\textwidth,keepaspectratio]{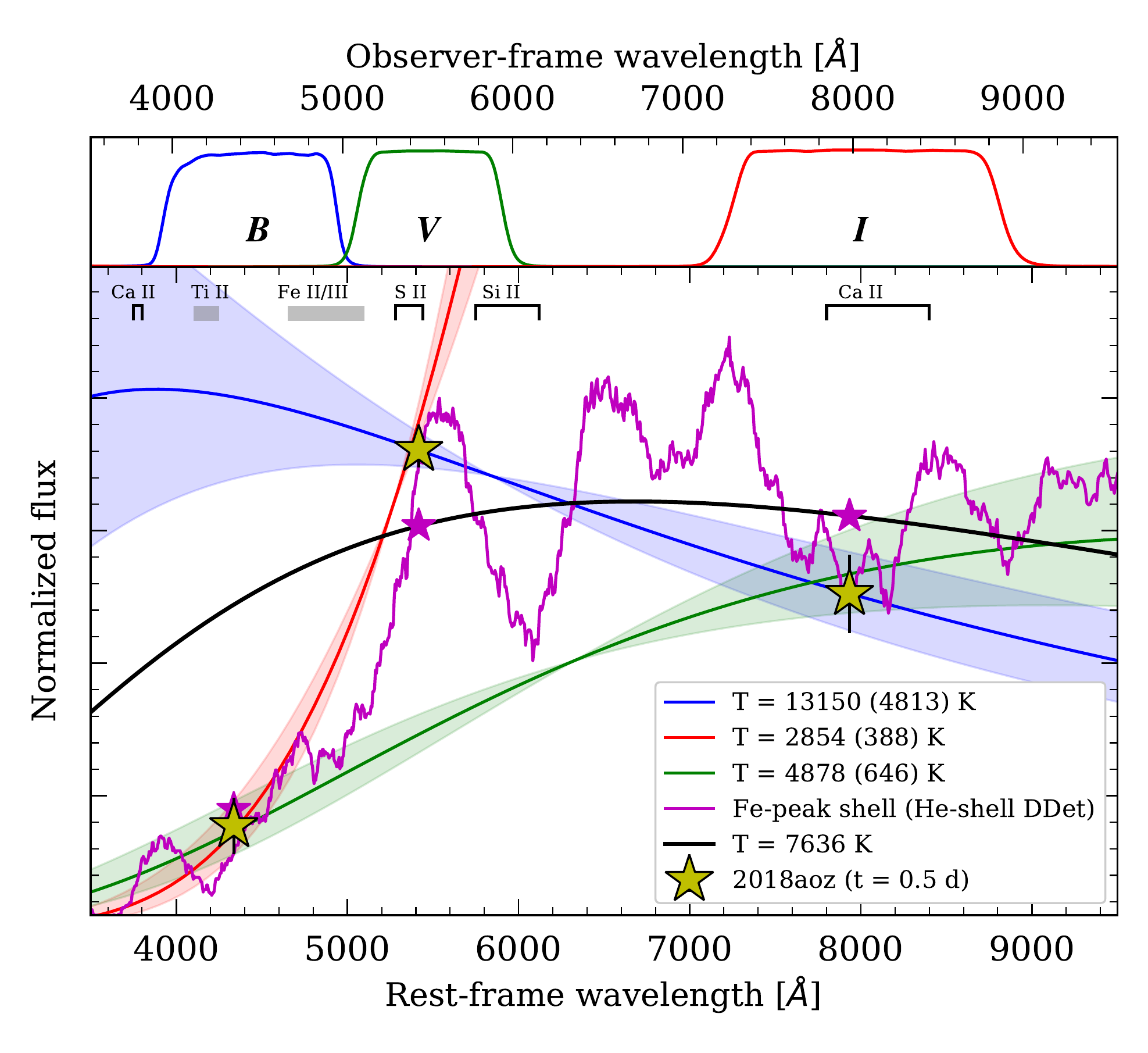}
\caption{\textbf{Infant-phase SED of \uname\ compared to model predictions.} 
(Top) The KMTNet $BVI$ transmission curves used for $BVi$ photometry.
(Bottom) The dereddened $BVi$ fluxes of \uname\ (yellow stars) at 0.5 days (0.9 days post-explosion), when the $B$-band suppression relative to the $Vi$-band flux is maximal, are compared to the following three cases: (1) blackbody distributions (colored curves with shaded 1-$\sigma$ error regions) fitting the $B-V$ (red), $V-i$ (blue), and $B-i$ (green) colors whose temperatures ($T$) are provided in the legend with 1-$\sigma$ uncertainties in parentheses; (2) simulated (magenta) spectrum of a 0.01$\,$\msol\ He-shell $+$ 1.05$\,$\msol\ WD He-shell DDet from 1.05 days post-explosion with Fe-peak elements in the outer 1\% of the SN-ejected mass (notable absorption features are labelled at the top of the panel); and (3) a blackbody (black curve) compatible with the $Vi$-band fluxes of the simulated He-shell DDet spectrum (magenta stars) evaluated using the KMTNet transmission curves.
The SN fluxes are from $>$ 7-$\sigma$ detections and their errorbars represent 1-$\sigma$ uncertainties.
The magenta spectrum shows how a shell of Fe-peak elements near the ejecta surface can lead to selective $B$-band suppression in the infant phase.
}
\label{fig:specfig}
\end{figure}

\clearpage

\setcounter{figure}{0}
\setcounter{table}{0}
\renewcommand{\figurename}{\textbf{Supplementary Figure}}
\renewcommand{\tablename}{\textbf{Supplementary Table}}

\section*{Supplementary Information}

\section{Power-law Fit for the Epoch of First Light.}

To estimate the epoch of first light of \uname, we use a power-law model whose form is $f_{\nu}\propto (t-t_{\rm efl})^{\alpha_{\nu}}$, where $t$ is time (days) since first light, $f_{\nu}$ is flux in filter $\nu$ normalized by peak flux, $t_{\rm efl}$ is the epoch of first light, and $\alpha_{\nu}$ is power-law index.
$\alpha_{\nu}$ is expected to be $\sim$ 2 for thermal emission from a homologously expanding \ni56-powered sphere\cite{Nugent2011nat}, though it can vary depending on the ejecta structure.
The $BVi$-band rising light curves of \uname\ up to $\sim$ 40\% of peak brightness\cite{Olling2015nat} (excluding that of the $B$-band during 0--0.5 days) were fitted with a model consisting of a single power-law in each band sharing a common $t_{\rm efl}$.
The best-fit model (Figure~3, left) estimates the epoch of first light at $23.9\pm0.5$~h on 2018 March 28 UT (MJD $58205.996\pm0.020$) and power-law indices of $2.24\pm0.02$, $1.99\pm0.02$, and $2.26\pm0.01$ in the $B$, $V$, and $i$ bands, respectively, similar to indices found in other \tase\cite{Miller2020apj, Dimitriadis2019apj, Olling2015nat}.
We note that while the power-law cannot be applied to the plateaued $B$-band data before 0.5 days, the non-plateaued $B$-band data after 0.5 days fits the power-law excellently, and, additionally, we find that excluding the latter data does not change the best-fit epoch of first light.
The power-law also provides an adequate fit to the $Vi$-band light curves of the SN outside of the infant phase where there is observed excess emission.
This is confirmed by examining the contribution to the total \chisqr\ error from different epochs, with \chisqr\ = 6.0 coming from the infant phase (0--1 days) and only \chisqr\ = 3.2 coming from all $BVi$ data in subsequent epochs (1--7 days).
We obtain upper and lower bounds for the epoch of first light when all three infant-phase light curves are included in and excluded from the fit, respectively. The best-fit model in each case is shown as the upper and lower bounds, respectively, of the grey shaded error region in Figure~3.
For these models, the fitted $t_{\rm efl}$ values are $0.08\pm0.03$ days after and $0.28\pm0.09$ days prior to  MJD $58205.996$, respectively.

\section{Photospheric Velocity and the Explosion Epoch.}

The photospheric velocities of \tas\ ejecta are typically measured using the Doppler shift of the Si~II~$\lambda$6355~\AA\ absorption feature (or ``Si~II velocity''), which persists until $\sim$ 30 days after peak\cite{Tanaka2008apj,Wang2009apj,Parrent2014apss}.
We obtain the Si~II velocity evolution of \uname\ by fitting the minimum of the 6355~\AA\ feature with a skewed Gaussian profile for a series of spectra obtained between $-$11.0 and 30.7 days since $B$-band maximum (Supplementary Figure~\ref{fig:specevol}, left panel).
Supplementary Figure~\ref{fig:specevol} (bottom-right panel) shows that the measured Si~II velocity evolution of \uname\ is consistent with the range of Si~II velocities observed in normal \tase\ 2011fe\cite{Pereira2013aa} and 2002dj\cite{Pignata2008mnras}.

We estimate the explosion epoch of \uname\ by fitting its Si~II velocity evolution with a power-law model of the form $v_{\rm Si}\propto (t-t_{\rm exp})^{-\beta}$, where $v_{\rm Si}$ is the velocity and $t-t_{\rm exp}$ is time (days) since the explosion epoch. 
Such a model has been found to be a good fit to the pre-peak $v_{\rm Si}$ evolution in other \tase\cite{Piro&Nakar2014apj}.
We fit the $v_{\rm Si}$ of \uname\ before $B$-band maximum using the model, adopting $\beta$ = 0.22, the theoretically expected value for \tase\ with polytropic (n=3) ejecta structure.
Supplementary Figure~\ref{fig:specevol} (bottom-right panel) shows that the model (magenta line) provides a good fit to the observed velocities with \chisqr\ = 2.3.
The best-fit $t_{\rm exp}$ using this model is 0.4 $\pm$ 0.7 days prior to first light (or MJD 58205.6 $\pm$ 0.7), consistent with what has been found in other events with $t_{\rm efl} - t_{\rm exp}$ in the range of $-$0.3 to 1.6 days\cite{Piro&Nakar2014apj}.
The quality of the fit does not change significantly for slight variations of $\beta$ in the range of 0.20--0.24, resulting in \chisqr\ in the range of 2--3 and variations of up to 0.7 days in the inferred $t_{\rm exp}$.
The weighted mean of $t_{\rm exp}$ using three values of $\beta$ = 0.20, 0.22, and 0.24 is 0.2 $\pm$ 0.4 days prior to first light, which is smaller than 0.4 $\pm$ 0.7 days obtained using the theoretically expected value of $\beta$ = 0.22.
Overall, these results point to an explosion epoch for \uname\ within $\sim$ 1 day of its epoch of first light.

\section{Spectral Energy Distribution and Bolometric Light Curve.}

We construct the spectral energy distribution (SED) of \uname\ based on four data sets (Supplementary Figure~\ref{fig:lcall}):
(1) high-cadence optical photometry from KMTNet, which provides simultaneous measurements in $BVi$ bands;
(2) photometry in a few additional optical bands from LCO, which provides nearly simultaneous high-cadence measurements showing consistent evolution with KMTNet in the $B$-, $V$- and $i$-bands (see Supplementary Figure~\ref{fig:lcall});
(3) UV photometry from Swift with nearly simultaneous measurements in $UVM2$, $UVW1$, $U$, $B$, and $V$ bands at low cadence; and
(4) NIR photometry from ANDICAM, which provides asynchronous $JHK$ band measurements at low cadence.
Note that $UVW2$ is not included since its effective wavelength can be redder than that of $UVM2$ for \tas\ spectra\cite{Brown2016aj}.
In order to construct the full SED of \uname\ from these asynchronous light curves, 
we used a Gaussian process\cite{george} with a stationary Matern-3/2 kernel to interpolate the light curve in each band. 

We estimate the rate of radioactive energy production in \uname\ by constructing the bolometric light curve as follows.
Following the onset of Swift observations at 10.7 days before peak, we obtain a UVO (UV and optical) lower bound on the bolometric luminosity by integrating the photometric SED from Swift $UVM2$ band (2246~\AA) to $i$ band (7698~\AA) using trapezoidal rule.
Since the majority of the emission from \tase\ falls within the UVOIR (UV, optical, and NIR) waveband during its photospheric phase\cite{Contardo2000aap}, roughly from $U$ band to $K$ band, we complete the bolometric luminosity by extending the trapezoidal integral to the $K$ band (23763~\AA) whenever ANDICAM photometry is available.
Where ANDICAM coverage is lacking, we substitute NIR luminosity from the best-fit SNooPy template (Methods), extrapolating the template out to $K$ band.
The difference between the UVO and UVOIR bolometric light curves caused by the NIR tail extrapolation accounts for $<$ 20\% of the total bolometric luminosity.
Supplementary Figure~\ref{fig:arnett} presents the SED bolometric light curve of \uname\ (black filled circles) alongside a bolometric light curve constructed by directly integrating the SNooPy template from 2246~\AA\ to 23763~\AA\ (black open circles), showing that the template bolometry represents the observed one reasonably well.
Thus, we extend the bolometric light curve of the SN with the integrated SNooPy template until $\sim$ 15 days before $t_p$ (or $\sim$ 1.3 days since first light), where the valid range of the template begins\cite{Hsiao2007apj}.

\section{Explosion Parameters.}

We estimate the \ni56\ mass, ejecta mass, and ejecta kinetic energy of \uname\
by comparing its bolometric light curve to the \ni56-powered SN light curve model by Arnett (1982)\cite{Arnett1982apj} which is based on the following four assumptions: (1) the SN ejecta is spherically symmetric, (2) the light curve is powered only by \ni56\ decay, (3) the distribution of \ni56\ is strongly peaked toward the center of the ejected mass, and (4) the opacity is constant.
(Note that the opacity of \tase\ in the photospheric phase is expected to be nearly constant\cite{Pinto&Eastman2000apj}).
In the model, the SN luminosity is mainly determined by two parameters: \ni56\ mass and the $\tau_m$ timescale related to the ejecta mass and kinetic energy by Equation~2\cite{Moon2021apj}.
We fit the model for two cases: one where the onset of the model, representing the beginning of the central \ni56-powered light curve, is equal to the epoch of first light (see Supplementary Section 2) and another where the onset is allowed to vary.

In the first case, the Arnett model is matched to the peak bolometric luminosity of \uname\ ($1.43\times10^{43}$~ergs~$^{-1}$) and the rise time of the light curve from the epoch of first light to the bolometric peak (14.17 days in the rest frame), as shown in Supplementary Figure~\ref{fig:arnett} (red dotted curve).
The inferred ejecta parameters are \ni56\ mass $=0.63\pm0.01$~\msol\ and $\tau_m = 12.9\pm0.2$ days.
However, as seen in the figure, this model poorly matches the observed bolometric light curve, whose evolution near the peak is much faster, and fails to represent the rapid rise of the SN.
In the second case, we fit the Arnett model to the bolometric light curve near peak (from $-$10 days to 10 days since peak), allowing the time of the onset of the model to be a fit parameter, $t_{\rm Ni}$, representing the onset of the central \ni56-powered light curve.
The best-fit model, with inferred \ni56\ mass of $0.509\pm0.006$~\msol, $\tau_m = 9.51\pm0.26$ days, and $t_{\rm Ni} = 2.42\pm0.16$ days after first light in the rest frame, is shown in Supplementary Figure~\ref{fig:arnett} (green dot-dashed curve). 
As seen in the figure, the model obtained in this method provides a much better comparison to the observed bolometric light curve near the peak (\chisqr\ $\sim$ 0.2), though it underpredicts luminosities during earlier epochs.
The discrepancy can be explained by a shallower \ni56\ distribution, which provides additional emission at earlier times (see below).

We estimate the ejecta mass and kinetic energy of \uname\ using the best-fit $\tau_m$ parameter ($9.51\pm0.26$ days) and Equation~2, adopting the observed Si~II photospheric velocity near peak as the characteristic ejecta velocity.
Using the approximate \ni56-dominated opacity expected for \tase\ during the photospheric phase ($\kappa\sim 0.1$~cm$^2$~g$^{-1}$)\cite{Pinto&Eastman2000apj}, we obtain fiducial values of $0.80\pm0.04$~\msol\ and $(0.63\pm0.04)\times 10^{51}$~ergs for the ejecta mass and kinetic energy, respectively.
By extending the calculation to include a range of opacities between $0.08$ and $0.10$~cm$^2$~g$^{-1}$ that have been used in modelling the bolometric luminosity of \tase\cite{Piro&Nakar2014apj, Arnett1982apj, Li2019apj}, we estimate the ejecta mass and kinetic energy of \uname\ to be in the ranges of $\sim$ 0.8--1.0~\msol\ and $\sim$ (0.6--0.8) $\times$ 10$^{51}$~ergs, respectively.
In a comparison of the \ni56\ and ejecta masses of \uname\ with what has been found in other normal \tase\cite{Scalzo2019mnras}, the \ni56\ mass of $0.509\pm0.006$~\msol\ is consistent with the expectations for normal \tase\ with similar \sbv\ as \uname, but the estimated ejecta mass range of 0.8--1.0~\msol\
is located at the lower extreme of what has been found in other normal events ($\sim$ 0.9--1.2~\msol\ for \sbv\ = 0.8) and well below the Chandrasekhar mass (1.4~\msol).
This shows that
an ejecta mass of \uname\ that is comparatively lower than other normal events can explain its relatively rapid pre-peak rise time (see Figure~1).
However, we note that the exact values of explosion parameters from such bolometric luminosity modelling in general can be subject to additional systematic uncertainties from the model assumptions.
For example, using radiative transfer simulations described in Supplementary Section 7.1 that account for \ni56\ stratification and time-varying opacity, we obtain a slightly higher ejecta mass of 1.05~\msol\ for \uname, though this method depends on the SN having an ejecta profile specific to the He-shell DDet explosion mechanism.
Similarly, for asymmetric explosion mechanisms, different ejecta masses can be obtained depending on viewing angle, and the mass inferred using 1-dimensional models should be interpreted as the ``effective'' ejecta mass along the line of sight rather than the total mass of the SN ejecta in such cases.

\section{Distribution of \ni56 Heating in the Ejecta.}

As discussed above, the Arnett model, while it provides useful measurements of the bulk properties of the ejecta, is inadequate in explaining the early bolometric light curves within $\sim$ 2--10 days post-explosion.
This may be caused by a \ni56\ distribution which is not strongly peaked toward the center of the ejected mass, and we explore whether heating from a shallower \ni56\ distribution, extended to the progenitor surface, can account for the early luminosity.
According to Piro and Nakar (2014)\cite{Piro&Nakar2014apj}, who investigated the impact of \ni56\ distribution on the early light curve, radioactive heating from a logistically stratified distribution of centrally concentrated \ni56\ is adequate in explaining the rising light curves of other normal \tase.
In their model, the fraction, X$_{56}$, of \ni56\ per mass of ejecta at depth $x$ in the progenitor, measured from the surface of the star, follows a logistic function:
\begin{equation}
	X_{56}(x) \propto \frac{1}{1+\exp{[-\beta(x-x_{1/2})]}} 
\label{eq:ni}
\end{equation}
\noindent
where the depth coordinate ($x$) is in units of the diffusion time ($t_{\rm diff}$), the time in days since explosion when the ejecta becomes optically thin; $\beta$ describes how fast the logistic distribution declines;
and $x_{1/2}$ is the depth where the distribution is at half-maximum. 
The normalization of Eqn~\ref{eq:ni} is determined by the observed luminosity of the SN at the diffusion time. 
Note that this model is only applicable to the light curves of SNe during early phases before the peak\cite{Piro&Nakar2013apj} where it has been used to investigate the stratification of \ni56\ in shallow layers of the ejecta\cite{Piro&Nakar2014apj}.

Adopting $\tau_m = 9.51$ from bolometric luminosity modelling (see Supplementary Section 4 above) and assuming
that the epoch of explosion is roughly equal to that of the first light, we fit the logistic \ni56\ model to the bolometric light curve during 2--10 days since first light.
Note that the difference between the epochs of explosion and first light in \uname\ is estimated to be $\lesssim$ 1 day and consistent with zero (see Supplementary Section 2).
Additionally, adopting $\lesssim$ 1-day earlier explosion epochs in the model does not affect the quality of the fit significantly since the slight delay before first light is simply explained by a slightly deeper \ni56\ distribution (i.e., larger $\beta$ and/or $x_{1/2}$), as detailed below.
Supplementary Figure~\ref{fig:arnett} shows the best-fit model (blue dashed curve) with $x_{1/2} = 0.31 \pm 0.04$, $\beta=8.5 \pm 1.0$, and $t_{\rm diff} = 20.8 \pm 0.4$~days.
As seen in the figure, the model provides an excellent match to the observed bolometric light curve of \uname\ within the range of 2--10 days, confirming that radioactive heating from a logistic distribution of \ni56\ can indeed explain the observed luminosity in that period.

We fitted the infant-phase $BVi$-band light curves of \uname\ using the logistic \ni56\ model by adopting a blackbody SED for the modelled emission.
In the model, the blackbody temperature of the SN emission 
($T_c$; also known as color temperature) is determined by 
$T_c^4 = L \tau_c /(4\pi \sigma_{SB} r_c^2)$,
in which $L$ is the luminosity and $\tau_c$ is the optical depth 
at the ``color depth'' ($r_c$) where the SN radioactive emission is thermalized.
For the time evolution of $r_c$, we adopt the equation from Piro and Nakar (2013) describing the evolution of the photospheric radius\cite{Piro&Nakar2013apj}, assuming the color depth is located close to the photosphere.
Since the color depth is determined by the diffusion depth separating the adiabatically expanding inner ejecta from the outer layers that cool via photon diffusion,
we assume $\tau_c$ is a fraction of the optical depth at the diffusion depth, $\tau_c \sim \alpha \times c/v$.
The ejecta velocity ($v$) is extrapolated from the power-law fit to the Si~II velocity of the SN (Supplementary Figure~\ref{fig:specevol}, bottom-right panel).
The $BVi$-band light curves of \uname\ during 0--10 days after first light, excluding the $B$-band light curves $\lesssim$ 0.5 days, were fitted with the blackbody logistic \ni56\ model, obtaining the best-fit $\alpha = 0.054$.
The best-fit model (Figure~3, blue dotted curves) adequately explains the overall luminosity
evolution of \uname.
However, as noted in section ``Distribution of Fe-peak Elements in the Extreme Outer Ejecta,'' the model underpredicts the observed light curves in all three $BVi$ bands during the infant phase (Figure~3, left panels), showing that a lower SN luminosity is expected during the phase if the emission comes from a logistic, or similarly smoothly stratified and monotonic, distribution of \ni56\ alone.

To investigate if additional radioactive heating from \ni56\
near the ejecta surface can explain this luminosity difference
during the infant phase, we modify Equation~\ref{eq:ni} to include
two new fit parameters separately representing the \ni56\ distribution near the surface: $x_s$ and $M_s$ for location and mass, respectively.
In the modification, the \ni56\ mass fraction in depths shallower than a fixed position,
or $X_{56}(x < x_s)$, is adopted to be constant such that the total integrated mass
of \ni56\ in the shell described by $x < x_s$ is equal to $M_s$.
We fit the model to the $V$-band light curve of \uname\ during 0--1 days after first light, 
obtaining the best-fit with $M_s = 1.8\times10^{-4}$~\msol\ and $x_s=0.015$, 
corresponding to a shell of \ni56\ in the outer 0.31\% of the ejected mass.
By varying the explosion epoch by up to $\sim$ 1 day before the epoch of first light, we find that the explosion epoch mainly affects $\beta$, increasing it up to $\sim 20$, while $x_{1/2}$ and $t_{\rm diff}$ both vary by $<$ 10\%, and the inferred mass of excess \ni56\ responsible for the early excess emission varies by $<$ 17\%.
The red, green, and purple models in Figure~3 (right panels) were obtained by adopting $M_s=(0.9$, $1.8$, and $3.6)\times10^{-4}$~\msol, respectively, while keeping the other best-fit parameters fixed.
As seen in the figure, a shell of \ni56\ near the ejecta surface can indeed explain excess luminosity during the infant phase over the expectation from the best-fit logistic \ni56\ distribution, which represents the main distribution of radioactive elements in the SN ejecta.

\section{Non-continuous Process for the $B$-band Suppression.}

As noted in section ``Origin of the Infant-Phase $B$-band Plateau and Red $B-V$ Color,'' no blackbody can fit the $BVi$ SED of \uname\ at 0.5 days since first light without a $B$-band suppression.
One conceivable origin of the blue flux suppression at these early phases is extinction by circumstellar dust near the progenitor star.
However, as detailed below, this process is incompatible with what is observed in \uname, which requires significantly non-continuous extinction to explain the $B$-band suppression. 
For this, we examine if blackbody $+$ extinction can reproduce the infant-phase $BVi$ SED of \uname.
We compare in Supplementary Figure~\ref{fig:ext} (top-left) the observed SED (colored symbols) at two epochs in 0--0.5 days to those of a set of blackbodies only (4000--14000~K; colored solid curves) and blackbodies $+$ extinction (colored dashed curves).
The 14000~K blackbody is consistent with the observed $V-i$ color at 0.5 days, 
while the 4000~K case is for the $B-V$ color.
(Note that the latter is an unreasonably low blackbody temperature for early Type Ia SNe).
We adopt $R_V$ = 3.1 based on Galactic interstellar extinction\cite{Fitzpatrick&Massa2007apj} and a range of $E(B-V)$ in 0--1 mag using the same blackbodies from the blackbody-only models.
As shown in the figure, cooling (14000 to 4000~K) blackbodies with $R_V$ = 3.1 extinction in the early phase $<$ 0.5 days could mimic the observed evolution of the $B-V$ color, with $E(B-V)$ = 0.5 mag as an exemplary case, but they cannot accommodate the much bluer $V-i$ color with the overall goodness of the fit to the $BVi$ bands deteriorating for increased extinction.

We further investigate whether blackbody $+$ extinction with a smaller value of $R_V$,
which effectively increases the relative extinction in $i$ band\cite{Fitzpatrick1999pasp},
between 0.1 and 3.1 can fit the observed 
$BVi$-band fluxes of \uname\ at 0.5 days, the epoch with the lowest $B$-band flux compared to the $V$ and $i$ bands.
As explained below, 
we find that no reasonable blackbody $+$ extinction model can accommodate 
the observed SED even with smaller $R_V$ values, 
although smaller $R_V$ values do provide an improved fit,
with $R_V$ = 0.2 being the best case. 
Supplementary Figure~\ref{fig:ext} (middle- and bottom-left panels) shows the relationship between $E(B-V)$, temperature ($T$), and goodness of fit ($\chi^2$) from fitting 
the $BVi$ SED of \uname\ at 0.5 days with blackbody $+$ extinction for 
three exemplary cases of $R_V$ = 0.2, 1.7, and 3.1.
$R_V\gtrsim$ 1.7 fits worse than a blackbody alone for $E(B-V)$ $>$ 0 mag regardless of temperature, while $R_V\lesssim$ 1.7 fits better for increasing temperature and $E(B-V)$. 
$R_V$ = 0.2 provides the minimal $\chi^2$ for any temperature, though we note that the best-fit temperature itself is unreasonably large ($>$ 10$^7$~K).
In Supplementary Figure~\ref{fig:ext} (top-left panel), we compare the observed $BVi$ SED of \uname\ 
at 0.5 days with blackbody $+$ extinction for $R_V$ = 0.2 and two expectable temperatures in \tase: 
(1) $T$ = 10000~K, which can be typically expected in \tase\ at $\sim$ 1 day post-explosion\cite{Piro2010apj}; and
(2) $T$ = 25000~K, an exceptionally high temperature that can only
be expected in extreme cases, such as the ejecta colliding with a red giant companion\cite{Kasen2010apj}.
Note that the latter case is incongruent with the absence of bright excess emission
in the early light curve of \uname\ during 1--5 days; consequently,
$T$ = 25000~K can be considered as an upper limit of the blackbody temperature.
Although the $T$ = 25000~K case fits better than the 10000~K case, even the extreme case fails to reproduce the observed SED with $R_V$ = 0.2 extinction, while other $R_V$ fit worse for the same $T$. This shows that continuous extinction by dust cannot explain the $B$-band suppression in \uname.

We explore a potential non-continuous extinction process (e.g., line-blanket absorption) for 
the origin of the observed $B$-band suppression in \uname\ 
by modelling a blackbody SED with a spectroscopic break between the $B$ and $V$ bands.
For this, we calculate the required suppression of blackbody emission bluewards of a spectrum break for a range of break positions between 4000 and 5100 \AA\ that can match the observed $B$-band flux of \uname\ at 0.5 days.
We adopt the blackbody temperature of 7000~K as a conservative lower limit for the
temperature of \tase\ at this early epoch\cite{Piro2010apj, Piro&Nakar2014apj, Zheng2017apj} and normalize the blackbody to the average of the observed $V$ and $i$ band fluxes at 0.5 days in order to compare to the observations.
The solid line in Supplementary Figure~\ref{fig:ext} (right panel) shows the
required suppression as a function of spectrum break position from $\sim$ 4300 to 5100~\AA\
for the 7000~K blackbody.
As seen in the figure, for the spectrum break positions $\sim$ 4700~\AA,
nearly 100\% flux suppression is required even for the 7000~K case,
showing that smaller wavelengths are incompatible with the observed $B$-band suppression
since only blackbodies with lower temperatures can accommodate the observed suppression.
For the spectrum break positions in the range of 4700--5100 \AA,
$\gtrsim$ 70\% flux suppression is required to reproduce the
observed $B$-band suppression for the 7000~K blackbody,
with higher temperatures requiring more suppression as denoted by the non-hatched area above the solid curve.
Note that 6000~K and 8000~K blackbodies, shown with dashed lines below and above the solid line, respectively, provide the result for 1000~K variations, with a difference of only $\sim$ 5\% from the 7000~K case.
It appears, therefore, that the observed $B$-band suppression in \uname\
can be explained by a non-continuous extinction process
that can provide $\gtrsim$ 70\% reduction of $B$-band flux
relative to the $V$ and $i$ bands
by a spectroscopic break position between $B$ and $V$ band (i.e., from 4700 to 5100~\AA).

\section{Spectroscopic Predictions for Surface Fe-Peak Elements.}

Fe-peak elements produced in \tas\ nucleosynthesis are associated with a blanket of absorption lines bluewards of $\sim$ 5000~\AA\ at early phases when they are still in the second and third ionization states.
Recent radiative transfer modelling of \tas\ ejecta have predicted that the presence of nucleosynthetic elements near the ejecta surface leads to redward color evolution during the early phases due to Fe-peak line-blanket absorption within a few days post-explosion\cite{Maeda2018apj, Polin2019apj}.
Here, we examine a suite of models of \tas\ ejecta profiles with various distributions of Fe-peak elements near the surface, which include radiative transfer calculations, in order to determine if the Fe-peak line-blanketing effect can explain the observed $B$-band suppression and redward color evolution in \uname\ during its infant phase.

\subsection{7.1 Spectroscopic Predictions for Surface Fe-Peak Elements: He-shell DDet Simulations.}

First, we perform a grid of new He-shell DDet simulations with a range of core carbon-oxygen WD masses (1.00--1.10$\,$\msol) and He-shell masses (0.010--0.012$\,$\msol) following Polin et al. (2019)\cite{Polin2019apj}.
These simulations include a full treatment of hydrodynamics and nucleosynthesis using an isotope network in order to obtain the SN ejecta profile at each time step before conducting radiative transfer calculations of their spectro-photometric observables.
Note that similar simulations have been adopted to explain the light curves and spectra of spectroscopically peculiar \tase\ such as MUSSES1604D\cite{Jiang2017nat} and SN~2018byg\cite{De2019apj}.
Compared to these previous simulations, our simulations have smaller He-shell masses and predict observable properties that are more consistent with normal \tase.

Supplementary Figure~\ref{fig:ddet} (left panels) presents the results of our He-shell DDet simulations where the top-left and bottom-left panels show a comparison to the observed evolution of the $B-V$ color and the average $BVi$ flux density (in AB mag) of \uname, respectively.
We adopt the explosion epoch of \uname\ from Si~II velocity fitting, $0.4$ days prior to first light (see Supplementary Section 2), for each of the modelled light curves in the comparison.
As seen in the figure, the 1.05$\,$\msol\ WD + 0.010$\,$\msol\ He-shell model (magenta curves) provides an excellent fit to both the early $B-V$ color evolution of \uname\ and its overall luminosity evolution near peak.
The top-right panel presents the best-fit model prediction for the initial distribution of Fe-peak and associated nucleosynthetic elements---including $^{40}$Ca, $^{44}$Ti, $^{48}$Cr, $^{52}$Fe, and $^{56}$Ni---in the SN ejecta, showing an over-density of them in the outer 1\% of the ejected mass and a logistic-like distribution of \ni56\ peaked towards the ejecta center.
The bottom-right panel compares the observed spectra of \uname\ from $-$11.0, $-$1.9, $+$10.9 days after first light to those predicted by the best-fit model, showing that the modelled Fe~II/III absorption feature from Fe-peak elements agrees well with the observations throughout the evolution of the SN.

As shown in Figure~4 (bottom panel), the predicted spectrum of the best-fit He-shell DDet model at 0.5 days after first light can explain the observed $B$-band suppression of \uname\ in the infant-phase.
We note, however, that the model underpredicts both the observed $Vi$-band color temperature at 0.5 days (see Figure~4, bottom panel) and $BVi$-band luminosity during the infant phase (see Supplementary Figure~\ref{fig:ddet}, bottom-left inset), indicating that an additional source of heating is required to account for the missing emission.
Future studies are required to investigate if a He-shell DDet model incorporating such a heat source can indeed fully reproduce the infant-phase light curves of \uname.
While some of our He-shell DDet models can produce additional infant-phase emission, both the brightness of this emission and the duration of the redward color evolution scale with the He-shell mass. 
In particular, the 1.05$\,$\msol\ WD + 0.0111$\,$\msol\ He-shell model (Supplementary Figure~\ref{fig:ddet}, brown curves) provides a good match to the observed luminosity of the SN over the infant phase, but even this small increase in He-shell mass delays the redward color evolution by 1.5 days, inconsistent with our observations from 2--4 days.

\subsection{7.2 Spectroscopic Predictions for Surface Fe-Peak Elements: Monotonic \ni56\ distribution.}

In the He-shell DDet models, the Fe-peak elements near the surface of the SN ejecta responsible for the infant-phase $B$-band suppression form an independent shell (see Supplementary Figure~\ref{fig:ddet}, top-right panel), but these Fe-peak elements may instead be part of the tail of the monotonic and smoothly stratified distribution of \ni56\ that forms the central distribution of Fe-peak elements in the SN.
To investigate if a smoothly stratified and monotonic distribution can also explain the infant-phase $B$-band suppression and redward color evolution observed in \uname, we examine the set of radiative transfer models of logistically-distributed \ni56\ in \tas\ ejecta from Magee and Maguire (2020)\cite{Magee2020aa}.
In these models, the ejecta \ni56\ mass fraction has the following form:
\begin{equation}
	X_{56}(m) = \frac{1}{1+\exp{[-s(m-M_{Ni})/M_{\odot}]}} 
\label{eq:nirad}
\end{equation}
where $m$ is the mass coordinate from the ejecta surface, $M_{Ni}$ is the total \ni56\ mass, and $s$ describes how fast the logistic distribution declines.
As above, we adopt the explosion epoch of \uname\ from Si~II velocity fitting, $0.4$ days prior to first light, for the modelled light curves.

The ejecta density profile for the logistic \ni56\ radiative transfer models from 
Magee and Maguire (2020)\cite{Magee2020aa} follows either double-power-law or exponential form.
In the models, the parameters
$s$, $M_{Ni}$, and ejecta kinetic energy ($E_{\rm ej}$) also vary in the ranges of
3--100, 0.4--0.8~\msol, and 0.5--2.2 $\times 10^{51}$~ergs, respectively,
while the ejecta mass is fixed to be 1.4~\msol.
We find that these logistic \ni56\ radiative transfer models 
provide a good fit to the early $BVi$-band light curves of \uname\ during 1--10 days since first light with $s=9.7$, $E_{\rm ej}$ in the range of $0.8-2.2\times 10^{51}$ ergs, and $M_{Ni}=0.6$ and $0.8$~\msol\ for double-power-law and exponential density profiles, respectively.
The best-fit model has $s=9.7$, $E_{\rm ej}=1.8\times 10^{51}$ ergs,
and a double-power-law density profile.
Since the fixed ejecta mass in the models is larger than what we found in \uname\ (see above), the $E_{\rm ej}$ and $M_{Ni}$ values required to fit the observations are also larger.
The best-fit model is shown as the black dashed curve in Figure~2, while the full range of predicted $B-V$ colors for $E_{\rm ej}$ in the range of $0.8-2.2\times 10^{51}$ ergs is shown as the grey shaded region.
As noted in section ``Distribution of Fe-peak Elements in the Extreme Outer Ejecta,'' these models predict bluer and more continuous color evolution than the He-shell DDet simulation due to the smooth distribution of heating from logistically-distributed \ni56\ in the ejecta.

\clearpage

\begin{table}
\centering
\caption{Photometric observations of \uname.}
\medskip
\begin{tabular}{@{}lcccccr}
\hline\hline
Date (MJD) & Filter & m$_{obs}$ &  1-$\sigma$ (detection) & 1-$\sigma$ (total)\footnotemark[1] & S/N & Facility \\
\hline
58206.03779  & $B$ &    21.573  &       0.312  &        0.439  &        3.5    & KMTNet  \\
58206.03913  & $V$ &    21.257  &       0.254  &        0.254  &        4.3    & KMTNet  \\
58206.11241  & $B$ &    21.692  &       0.195  &        0.272  &        5.6    & KMTNet  \\
58206.11378  & $V$ &    20.686  &       0.090  &        0.091  &        12.1   & KMTNet  \\
58206.11528  & $i$ &    21.224  &       0.200  &        0.202  &        5.4    & KMTNet  \\
58206.17685  & $B$ &    21.768  &       0.190  &        0.265  &        5.7    & KMTNet  \\
58206.17834  & $V$ &    20.613  &       0.072  &        0.072  &        15.1   & KMTNet  \\
58206.17984  & $i$ &    21.154  &       0.162  &        0.164  &        6.7    & KMTNet  \\
58206.43054  & $V$ &    19.905  &       0.095  &        0.095  &        11.5   & KMTNet  \\
58206.43198  & $i$ &    20.074  &       0.162  &        0.161  &        6.7    & KMTNet  \\
58206.48679  & $B$ &    21.510  &       0.239  &        0.332  &        4.5    & KMTNet  \\
58206.48818  & $V$ &    20.029  &       0.087  &        0.088  &        12.5   & KMTNet  \\
58206.48953  & $i$ &    20.307  &       0.154  &        0.155  &        7.0    & KMTNet  \\
58206.49089  & $V$ &    20.086  &       0.094  &        0.094  &        11.6   & KMTNet  \\
58206.49238  & $i$ &    20.246  &       0.135  &        0.136  &        8.0    & KMTNet  \\
58206.54919  & $B$ &    21.566  &       0.210  &        0.292  &        5.2    & KMTNet  \\
58206.55053  & $V$ &    19.922  &       0.068  &        0.069  &        15.9   & KMTNet  \\
58206.55205  & $i$ &    20.121  &       0.141  &        0.142  &        7.7    & KMTNet  \\
58207.13608  & $r$ &    18.117  &       0.191  &        0.195  &        5.7    & DLT40   \\
58208.04432  & $B$ &    17.013  &       0.022  &        0.054  &        50.0   & KMTNet  \\
58208.04573  & $V$ &    16.529  &       0.019  &        0.025  &        55.8   & KMTNet  \\
58208.04728  & $i$ &    16.968  &       0.017  &        0.018  &        65.8   & KMTNet  \\
58208.12303  & $B$ &    16.882  &       0.016  &        0.048  &        68.5   & KMTNet  \\
58208.12449  & $V$ &    16.472  &       0.015  &        0.020  &        71.3   & KMTNet  \\
58208.12602  & $i$ &    16.911  &       0.015  &        0.016  &        71.7   & KMTNet  \\
58208.13576  & $r$ &    16.339  &       0.047  &        0.054  &        23.1   & DLT40   \\
58208.19667  & $B$ &    16.798  &       0.014  &        0.047  &        75.0   & KMTNet  \\
58208.19818  & $V$ &    16.416  &       0.014  &        0.019  &        75.9   & KMTNet  \\
58208.19971  & $i$ &    16.842  &       0.012  &        0.013  &        88.4   & KMTNet  \\
58208.45245  & $B$ &    16.580  &       0.021  &        0.052  &        52.3   & KMTNet  \\
58208.45389  & $V$ &    16.129  &       0.021  &        0.027  &        52.1   & KMTNet  \\
58208.45524  & $i$ &    16.585  &       0.015  &        0.017  &        70.3   & KMTNet  \\
\hline\end{tabular}
\begin{tablenotes}
\item {\bf Note.} The entirety of this table is available in machine readable form. Photometric data in each filter is calibrated according to Methods section ``Photometric Calibration.''
\item \footnotemark[1] Total 1-$\sigma$ error includes detection S/N, photometric calibration error, and S-correction error.
\end{tablenotes}
\label{tab:phot}
\end{table}

\begin{table}
\centering
\caption{Spectroscopic observations of \uname.}
\medskip
\begin{tabular}{@{}lcccr}
\hline\hline
Date (UT) & Telescope & Instrument & R & Wavelength [$\AA$] \\
\hline
2018 April 2.39 & COJ 2m & FLOYDS & 400--700 & 3250--10000 \\
2018 April 2.72 & COJ 2m & FLOYDS & 400--700 & 3300--10000 \\
2018 April 5.67 & COJ 2m & FLOYDS & 400--700 & 3250--10000 \\
2018 April 8.63 & COJ 2m & FLOYDS & 400--700 & 3250--10000 \\
2018 April 11.55 & COJ 2m & FLOYDS & 400--700 & 3250--10000 \\
2018 April 15.34 & MMT & Blue Channel & 3340 & 5700--7000 \\
2018 April 15.52 & COJ 2m & FLOYDS &  400--700 & 3350--9300 \\
2018 April 19.06 & du Pont & WFCCD & 800 & 3600--9200\\
2018 April 21.31 & du Pont & WFCCD & 800 & 3600--9200\\
2018 April 24.23 & MMT & Blue Channel & 3340 & 5700--7000\\
2018 April 24.35 & OGG 2m & FLOYDS & 400--700 & 3250--10000\\
2018 April 30.53 & COJ 2m & FLOYDS & 400--700 & 4800--10000\\
2018 May 4.57 & COJ 2m & FLOYDS & 400--700 & 3500--10000\\
2018 May 10.30 & OGG 2m & FLOYDS & 400--700 & 3500--10000\\
2018 May 13.08 & du Pont & WFCCD & 800 & 3600--9200\\
2018 May 14.25 & OGG 2m & FLOYDS & 400--700 & 3500--10000\\
2018 May 18.39 & COJ 2m & FLOYDS & 400--700 & 4800--10000\\
2018 May 23.46 & COJ 2m & FLOYDS & 400--700 & 4800--10000\\
2018 June 7.04 & du Pont & WFCCD & 800 & 3600--9200\\
2018 July 1.34 & COJ 2m & FLOYDS & 400--700 & 3500--10000\\
2018 July 6.96 & Gemini S & GMOS & 1690 & 5000--10000\\
2018 July 11.37 & COJ 2m & FLOYDS & 400--700 & 3500--10000\\
2018 June 14.01 & du Pont & WFCCD & 800 & 3600--9200\\
2018 July 20.39 & COJ 2m & FLOYDS & 400--700 & 3500--10000 \\
2018 July 22.35 & COJ 2m & FLOYDS & 400--700 & 3500--10000 \\
2018 July 24.35 & COJ 2m & FLOYDS & 400--700 & 3500--10000 \\
2018 August 12.99 & du Pont & WFCCD & 800 & 3600--9200\\
\hline
\end{tabular}
\begin{tablenotes}
\item {\bf Note.} The COJ 2m and OGG 2m are the Las Cumbres 2m Telescopes at Siding Spring Observatory and Haleakala Observatory, respectively. 
\end{tablenotes}
\label{tab:spec}
\end{table}

\begin{sidewaysfigure}
\includegraphics[width=\linewidth]{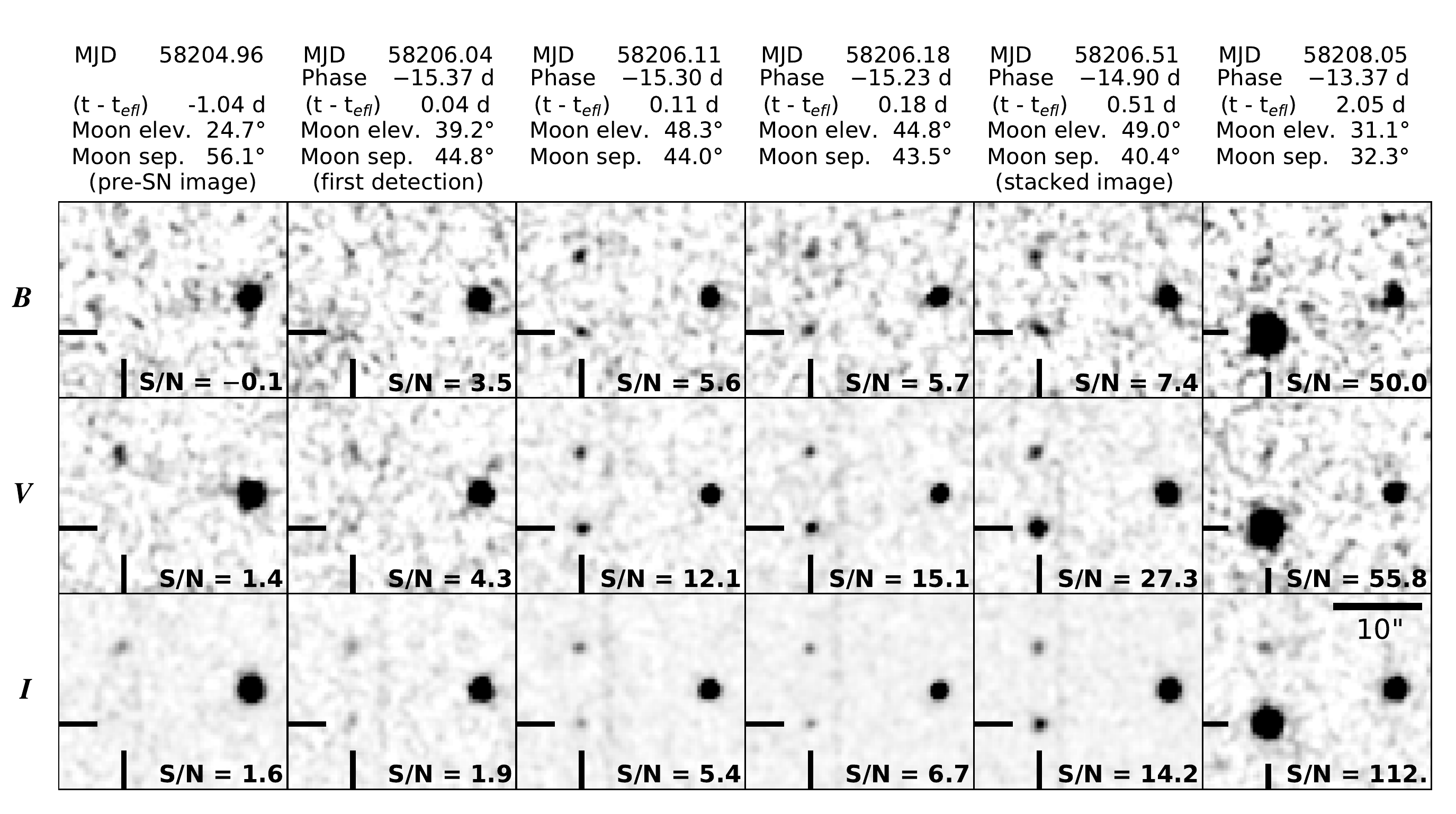}
\caption{\textbf{KSP images showing the first few detections of infant \uname.} 
KSP $BVI$-band images (rows 2--4) centered on the SN position in the epochs $\lesssim$ 2 days since the epoch of first light, showing the SN detections during the infant phase between 0 and 0.51 days (columns 2--5) as well as images taken immediately before and after the phase (columns 1 and 6). North is up and East is to the left.
The SN position is marked by crosshairs, and S/N is the signal-to-noise ratio of the source.
Note that the $I$-band S/N excludes the background source located 0.8\farcs\ from the SN position.
The 5th column images were obtained by binning 2--3 adjacent images within 1.5 hours of each other. The figure excludes one set of $BVI$ images from MJD 58206.43 due to low S/N.
The top table (row 1) shows the epoch corresponding to each column in MJD, days since $B$-band maximum (Phase), and days since the epoch of first light ($t-t_{\rm efl}$), as well as the moon elevation and angular separation from the source at each epoch.
The epochs corresponding to the $BVI$ images in the same column are within 0.01 days from each other.
}
\label{fig:det}
\end{sidewaysfigure}

\begin{figure}
\includegraphics[width=\linewidth]{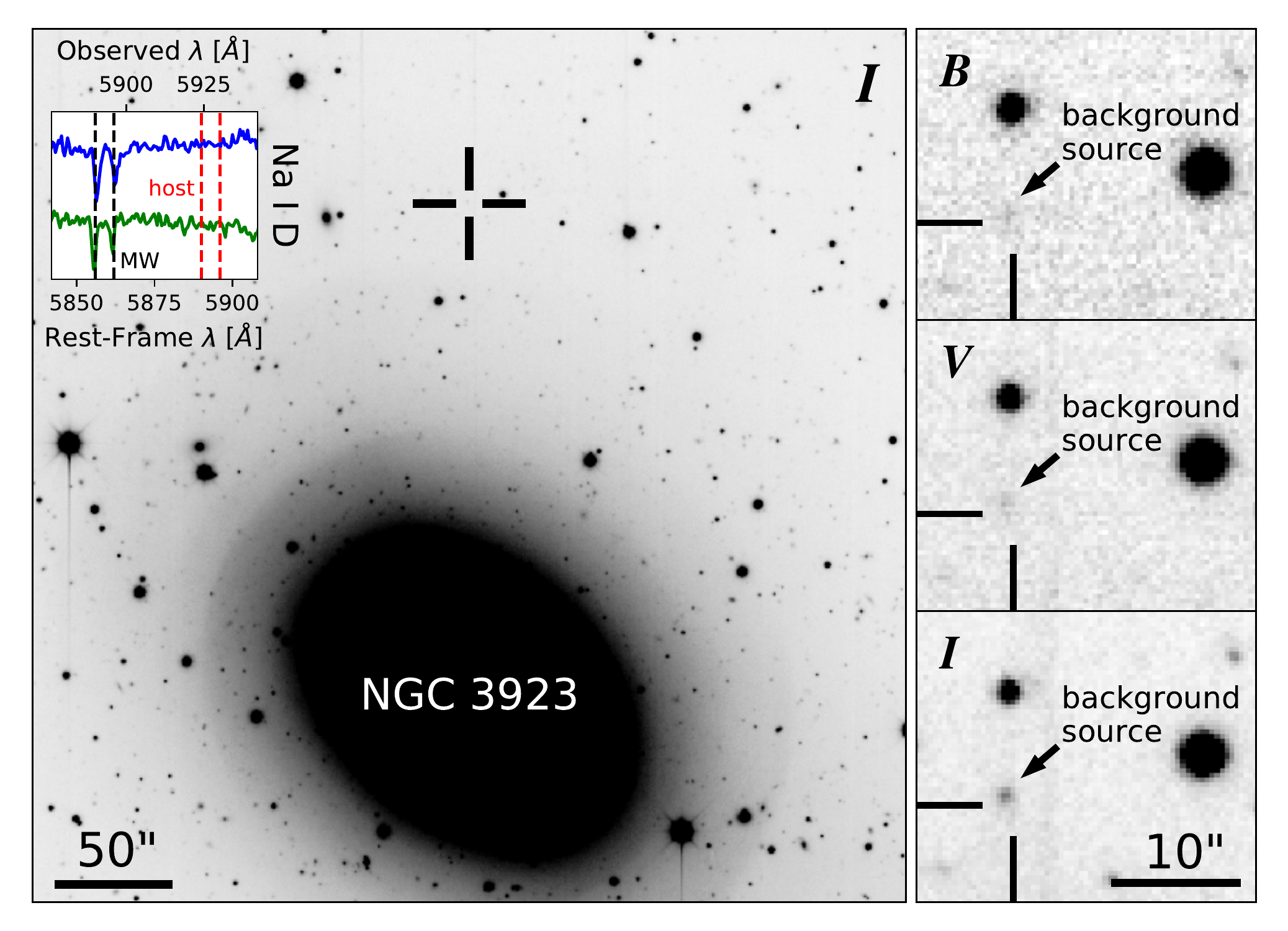}
\caption{\textbf{KSP stacked pre-SN images containing the source position.} 
(Left) KSP SNe~deep $I$-band stack of 336 images of the field containing the SN position (crosshair)
and the host galaxy NGC 3923 obtained before the first detection of the SN. North is up and East is to the left.
The inset shows moderate resolution MMT spectra of \uname\ from 1.9 (blue) and 10.8 (green)  rest-frame days since $B$-band maximum zoomed in on the detected Na I D feature at 5890 and 5896 \AA\ from the Milky Way (black vertical dotted lines) and the non-detection from the host galaxy (red vertical dotted lines).
(Right) KSP deep $B$-, $V$-, and $I$-band (top to bottom) stacks of 300, 313, and 336, images, respectively, containing the SN position (crosshair).
The arrow-indicated source, detected in all three stacked images $\sim$ 0.8\farcs\ offset north-east from the SN position, is a faint quiescent background star or globular cluster.}
\label{fig:host}
\end{figure}

\begin{sidewaysfigure}
\includegraphics[width=\linewidth]{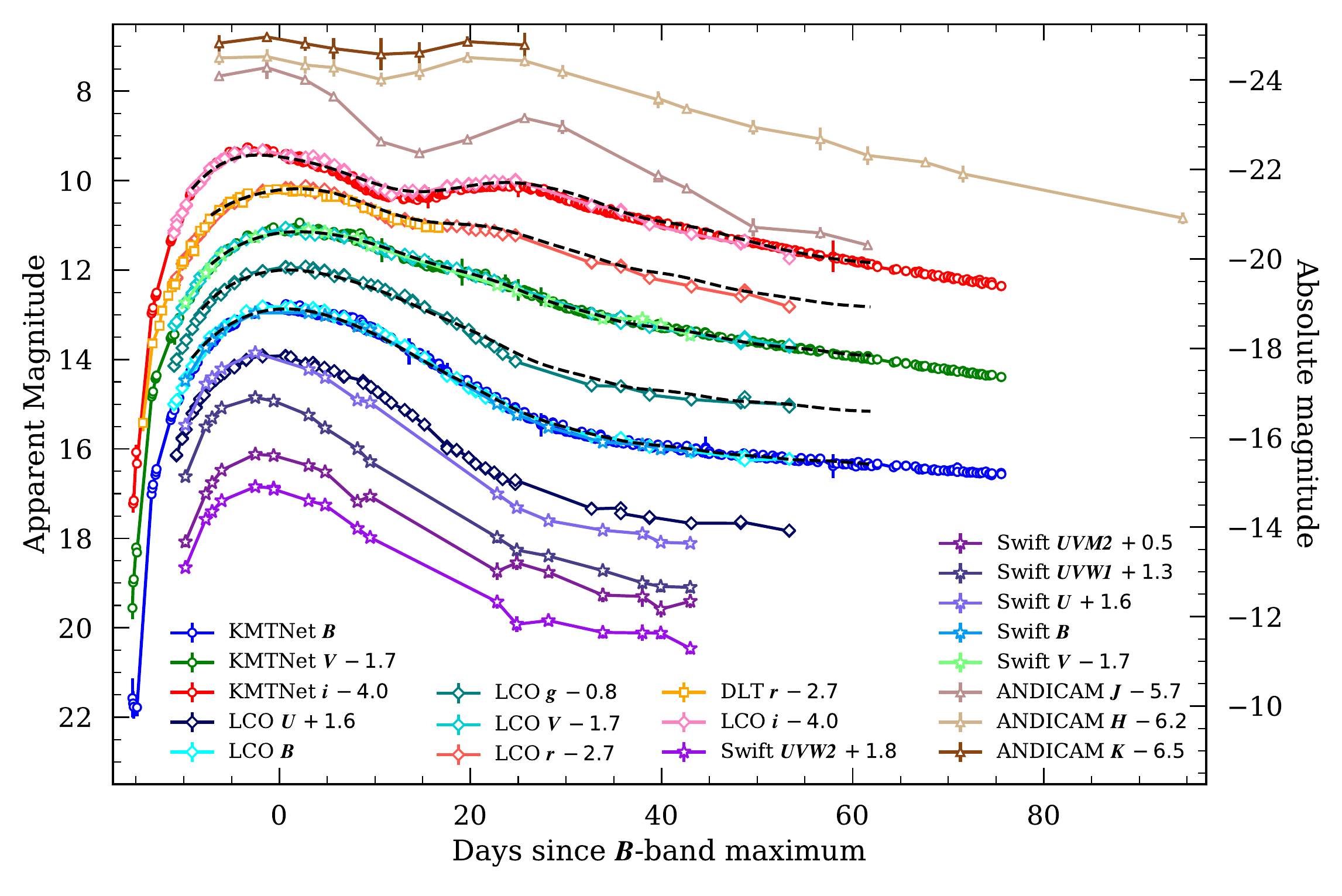}
\caption{\textbf{Observed multi-band light curves of \uname.} 
The optical light curves from KSP (circles), DLT40 (squares), and LCO (diamonds), UV light curves from Swift (stars), and the light curves from ANDICAM (triangles) are presented in the observer frame.
The light curves are vertically shifted for display clarity by the amount shown in the legend and the errorbars represent the 1-$\sigma$ uncertainty level. The black dashed curves compares the best-fit \tas\ template obtained using SNooPy (see Methods) to the observed $BgVri$-band light curves.
}
\label{fig:lcall}
\end{sidewaysfigure}

\begin{figure}
\includegraphics[width=\textwidth]{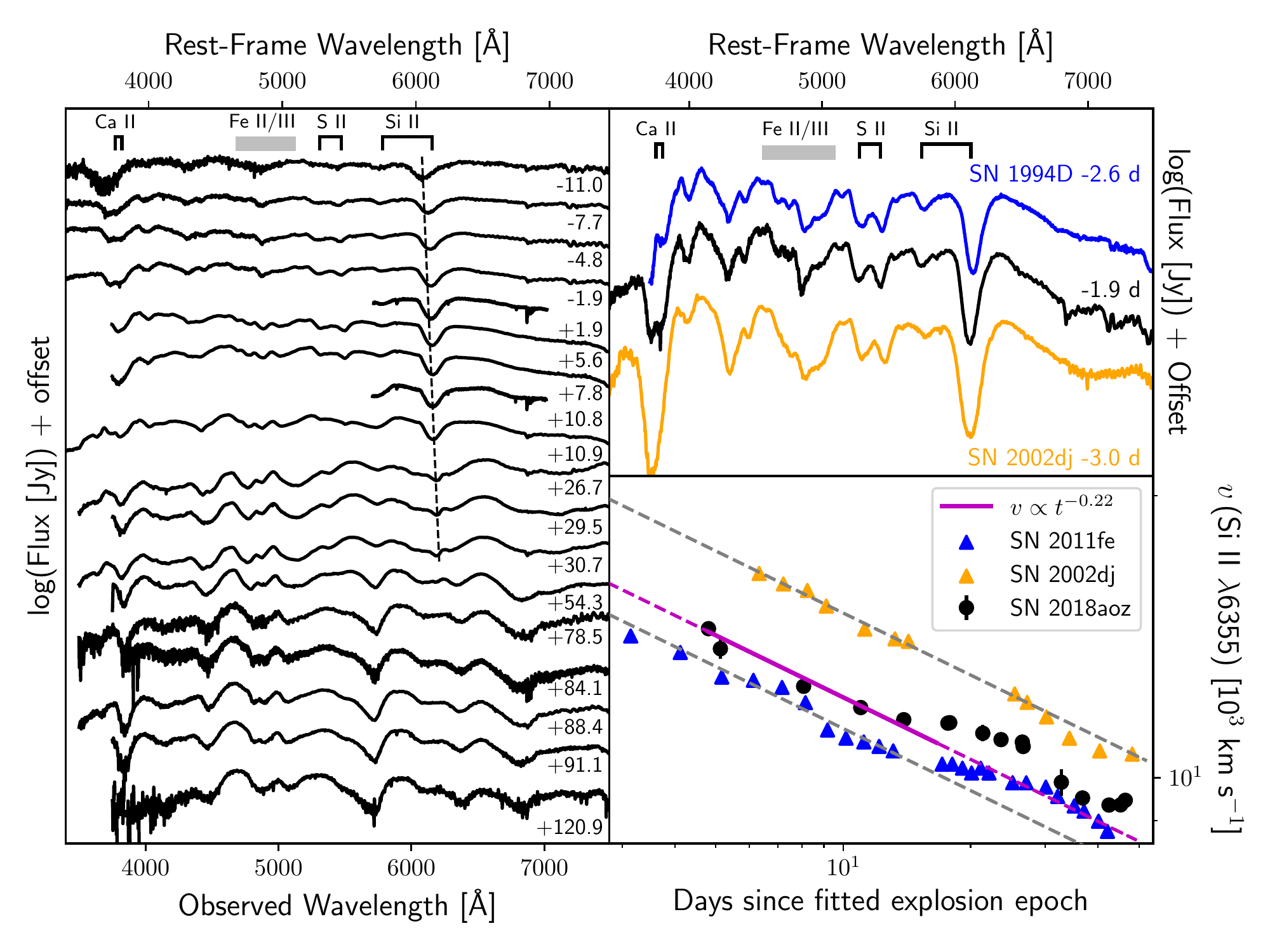}
\caption{\textbf{Spectroscopic observations of \uname.} (Left) A selection of medium-resolution spectra of \uname\ taken at various phases of the SN evolution, as given on the right side of the panel in rest-frame days since $B$-band maximum. The spectra were dereddened for Galactic extinction and offset vertically to avoid overlapping. Notable observed absorption features are labelled at the top of the panel. The dashed line traces the evolution of the Si~II~$\lambda$6355~\AA\ absorption feature. 
(Top-right) The spectrum of \uname\ (black solid line) taken near $B$-band maximum (1.9 days before) is 
compared to those of other normal \tase\ obtained at comparable epochs:
SNe~1994D\cite{Meikle1996mnras} (blue) and
2002dj\cite{Pignata2008mnras} (orange).
(Bottom-right) The rest-frame photospheric velocity evolution of \uname\ traced by the minimum of the Si~II~$\lambda$6355~\AA\ absorption feature is compared to the best-fit $v\propto t^{-0.22}$ power-law model (magenta; the solid part is the fitting interval) and two other normal \tase: 2011fe\cite{Pereira2013aa} and 2002dj\cite{Pignata2008mnras}.
The errorbars of the data points represent the 1-$\sigma$ uncertainty level.
}
\label{fig:specevol}
\end{figure}

\begin{figure}
\includegraphics[width=\textwidth]{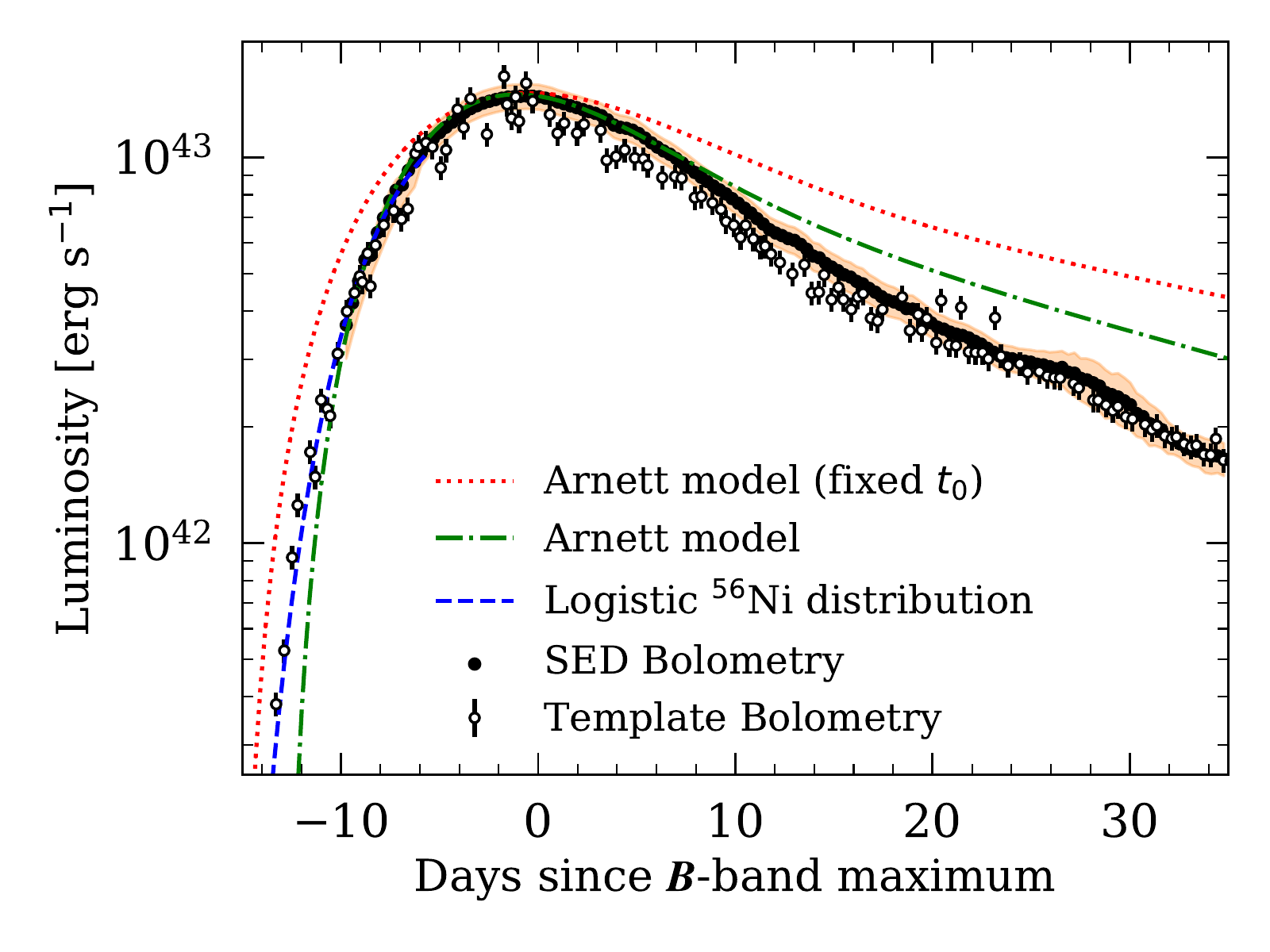}
\caption{\textbf{Bolometric light curve of \uname.} The bolometric light curve (black filled circles) of \uname\ constructed using its UVOIR SED is compared with the bolometric light curve of the SNooPy template (black open circles) and three models: (1) Arnett model\cite{Arnett1982apj} matched to the rise time and peak bolometric luminosity of \uname\ (red dotted curve), (2) Arnett model fitted to the bolometric light curve near peak (green dot-dashed curve), and (3) logistic \ni56\ model\cite{Piro&Nakar2014apj} (blue dashed curve).
The top of the pumpkin colored region corresponds to the 1-$\sigma$ uncertainty level of the SED bolometric light curve, while the bottom corresponds to the larger of the 1-$\sigma$ uncertainty level and the potential uncertainty from NIR template substitutions in the SED.
The errorbars of the black open circles represent 1-$\sigma$ uncertainties of the template bolometric light curve.
}
\label{fig:arnett}
\end{figure}

\begin{sidewaysfigure}
\includegraphics[width=\textwidth]{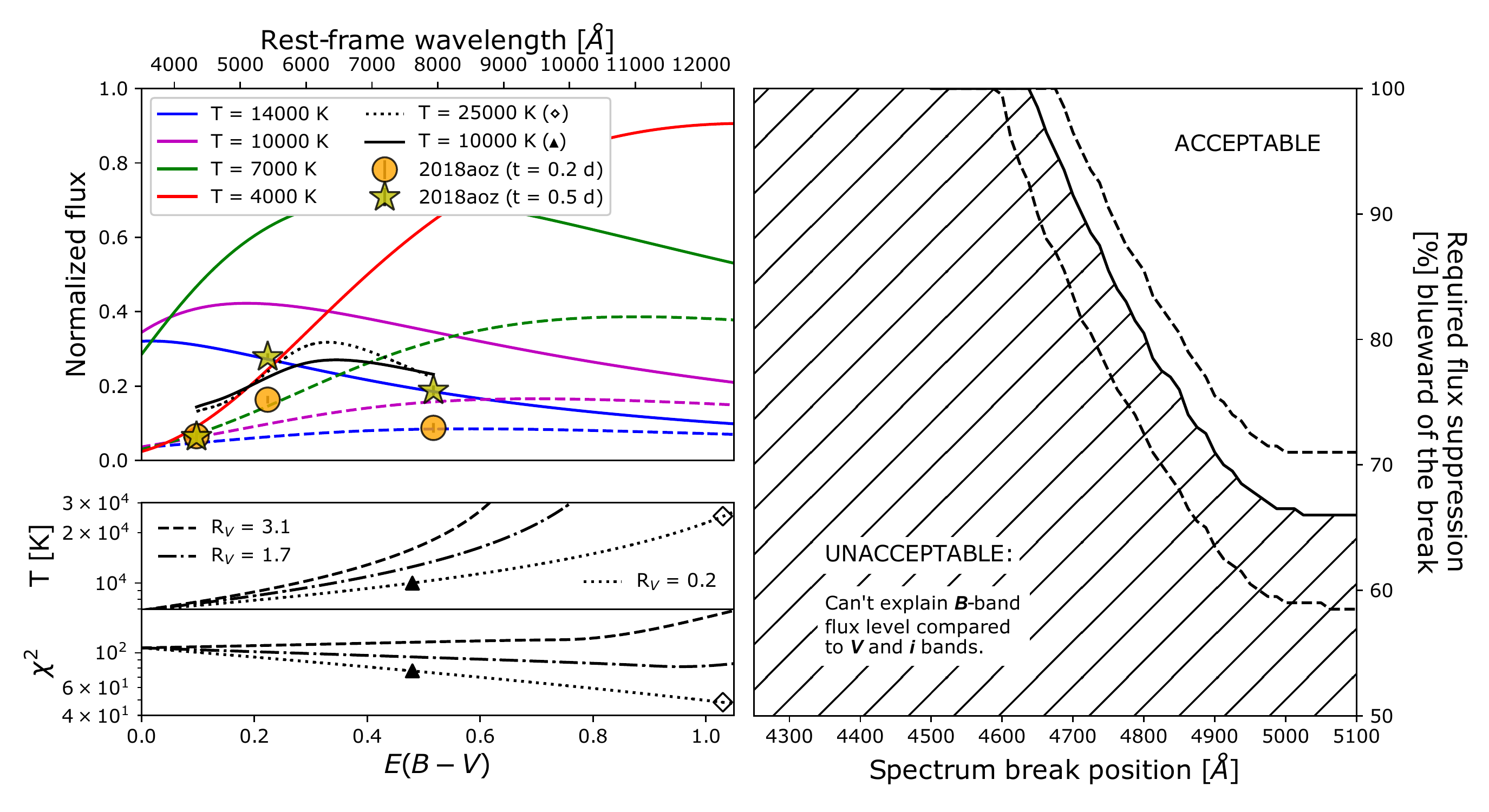}
\caption{\textbf{Comparison of the infant-phase SED of \uname\ to blackbody $+$ $B$-band suppression models.} 
\textbf{\textit{(Top-left)}} The $BVi$-band SED of \uname\ at 0.2 (orange circles) and 0.5 (yellow stars) days since first light, dereddened for Galactic extinction, are compared to the following models: (1) pure blackbodies at various temperatures between 4000~K and 14000~K ($T$; colored solid curves); (2) same as (1) but with added dust extinction of $E(B-V)$ = 0.5 with $R_V$ = 3.1 (same colors with dashed curves); and (3) fitted blackbodies ($T$ = 10000~K for black solid curve and 25000~K for black dotted curve) with $R_V$ = 0.2 extinction. 
\textbf{\textit{(Middle-left)}} The relationship between $T$ and $E(B-V)$ from fitting blackbody $+$ extinction to the observed SED of the SN at 0.5 days for $R_V$ = 0.2, 1.7, and 3.1.
The filled triangle and open diamond represent the 10000~K and 25000~K blackbodies from the top-left panel, respectively.
\textbf{\textit{(Bottom-left)}} The distribution of goodness of fit, $\chi^2$, over $E(B-V)$ from fitting blackbody $+$ extinction.
\textbf{\textit{(Right)}} 
The curves show the suppression factor (y-axis) required to reproduce the $B$-band flux level of \uname\ at 0.5 days using a blackbody normalized to the observed flux in the $V$ and $i$ bands with a spectroscopic break in the $B$ band (x-axis; flux bluewards of the break is suppressed).
The solid curve shows the required suppression for a 7000~K blackbody (solid curve), representing the minimum expectable temperature for infant \tase, and the dashed curves show those for 6000~K (below) and 8000~K (above) blackbodies.
The models in the hatched area provide insufficient suppression.
}
\label{fig:ext}
\end{sidewaysfigure}

\begin{figure}
\includegraphics[width=0.95\textwidth]{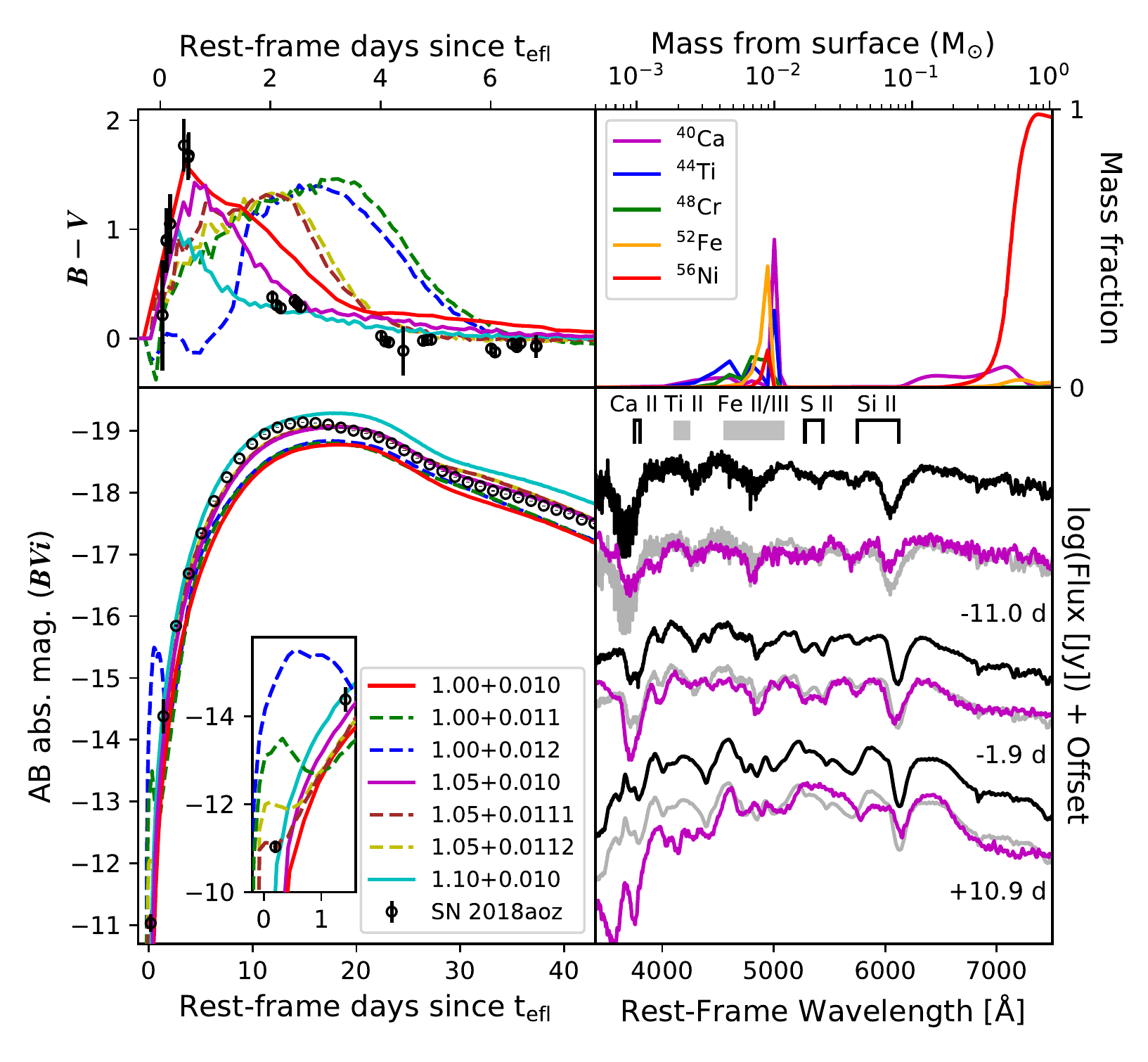}
\caption{\textbf{Outcome of He-shell DDet simulations and comparison to observations of \uname.}
(Left) Dereddened observations of \uname\ (black circles), including early $B-V$ color (top left) and 1-day binned $BVi$-averaged flux density (in AB mag; bottom left) evolution, are compared to the outcomes of He-shell DDet simulations (colored curves) for WDs and He-shells with varying masses as given in the bottom-left panel legend (WD mass + He-shell mass in \msol).
The errorbars of the data points represent the 1-$\sigma$ uncertainty level.
(Top right) The distribution of Ca, Ti, and Fe-peak elements (Cr, Fe, Ni) mass fractions in the ejecta from the best-fit model ($1.05+0.010$) at 4.6 seconds post-explosion. The x-axis is radial position ranging in 0--1.05$\,$\msol. 
(Bottom right) Three dereddened spectra (black) of \uname\ obtained from $-$11.0, $-$1.9, and 10.9 rest-frame days since $B$-band maximum.
The wavelengths of spectral features from Ca, Ti, S, Si, and Fe-peak elements expected in \tase\ are labelled at the top of the plot.
These observed spectra are translated downwards (grey) by subtracting a constant value for effective comparison with the best-fit He-shell DDet model predictions (magenta).
}
\label{fig:ddet}
\end{figure}

\newpage
\clearpage


\end{document}